\documentclass[12pt,twoside]{article}

\usepackage{jeep}       
\mysection{section}{\large \bf}{\thesection.~}   
\mysection{subsection}{\normalsize\bf}{\thesubsection.~}

\usepackage{a4wide}     

\usepackage{fancyheadings}

\usepackage{cite}   
\usepackage{epsf}
\usepackage{rotate}

\advance \headheight by 3.0truept       

\usepackage{times}

\newcommand{\lt}{\left}
\newcommand{\rt}{\right}
\newcommand{\ov}{\overline}
\newcommand{\rnc}{r_{ c \hspace{-1.1truemm}/ }}
\newcommand{\rqq}[3]{r_{#1 \ov{#2} #3}}
\newcommand{\eq}[1]{(\ref{#1})}
\newcommand{\as}{\ensuremath{\alpha_s}}
\def\openone{\leavevmode\hbox{\small1\kern-3.8pt\normalsize1}}%
\newcommand{\tw}{\ensuremath{\widetilde{\openone}}}
\newcommand{\1}{\ensuremath{\openone}}
\newcommand{\g}{\gamma }
\newcommand{\nn}{\nonumber \\}
\newcommand{\no}{\nonumber }
\newcommand{\fig}[1]{Fig.~\ref{#1}}
\newcommand{\tab}[1]{Tab.~\ref{#1}}
\newcommand{\bra}[1]{\langle \, #1 \, | }
\newcommand{\ket}[1]{| \, #1 \, \rangle }

\newcommand{\real}{{\rm Re}\,}

\newcommand{\prl}{Phys.~Rev. Lett.}
\newcommand{\prd}{Phys.~Rev.~D}
\newcommand{\plb}{Phys.~Lett.~B}
\newcommand{\npb}{Nucl.~Phys.~B}
\newcommand{\zpc}{Z.~Phys.~C}

\newlength{\miniwidth}
\newlength{\miniwidthplot}
\newlength{\nseparation}
\newenvironment{nfigure}[1]
        {\begin{figure}[#1]\hrule\vspace{\nseparation}\par}
        {\vspace{\nseparation}\par \hrule \end{figure}}
\newenvironment{ntable}[1]
        {\begin{table}[#1]\hrule\vspace{\nseparation}\par}
        {\vspace{\nseparation}\par \hrule \end{table}}

\begin{document}
~\\[-10truemm]
MPI-Ph/97-015 \hfill DESY 97-119 \\
TUM-HEP-274/97 \hfill hep-ph/9706501 \\
~\vspace{2.43truecm}
\begin{center}
{\LARGE Penguin Diagrams, Charmless B-Decays and the\\ 
  ``Missing Charm Puzzle''\footnote{Work supported by BMBF 
     under contract no.~06-TM-874.}}\\[3\baselineskip]
\textsl{Alexander Lenz\footnote{e-mail:alenz@MPPMU.MPG.DE},\\
Max-Planck-Institut f\"ur Physik --- Werner-Heisenberg-Institut,\\
F\"ohringer Ring 6, D-80805 M\"unchen, Germany, \\[2mm]  
Ulrich Nierste\footnote{e-mail:nierste@mail.desy.de},\\
DESY - Theory group, Notkestrasse 85, D-22607 Hamburg, Germany,\\[2mm]
and \\[2mm]
Gaby
Ostermaier\footnote{e-mail:Gaby.Ostermaier@feynman.t30.physik.tu-muenchen.de}\\
Physik-Department, TU M\"unchen, D-85747 Garching, Germany}
\end{center}
\vfill
\begin{center}
\textbf{\large Abstract} 
\end{center}
We calculate the contributions of penguin diagrams with internal $u$
or $c$ quarks to various inclusive charmless B-decay rates.  Further
we analyze the influence of the chromomagnetic dipole operator $Q_8$
on these rates. We find that the rates corresponding to
$\ov{B}\rightarrow X_{u \ov{u} s}$, $\ov{B}\rightarrow X_{d \ov{d}
  s}$, $\ov{B}\rightarrow X_{s \ov{s} s}$, $\ov{B}\rightarrow X_{s
  \ov{s} d}$ and $\ov{B}\rightarrow X_{d \ov{d} d}$ are dominated by
the new penguin contributions. The contributions of $Q_8$ sizably
diminish these rates. Despite of an increase of the total charmless
decay rate by 36 \% the new contributions are not large enough to
explain the charm deficit observed by ARGUS and CLEO. We predict
$n_c=1.33\pm 0.06$ for the average number of charmed particles per
B-decay in the Standard Model.  Then the hypothesis of an enhancement
of the chromomagnetic dipole coefficient $C_8$ by new physics
contributions is analyzed. We perform a model
independent fit of $C_8$ to the experimental data. If the CKM
structure  of the new physics contribution is the same as in the
Standard Model, $|C_8(M_W)|$ must be enhanced by a factor of 9 to 16
in order to explain the observed charm deficit.
\thispagestyle{empty}
\newpage
\section{Introduction}
Precision measurements performed at the $\Upsilon (4 S)$ resonance
find less charmed particles in the final states of B meson decays than
theoretically expected.  The CLEO 1.5, CLEO II and ARGUS data give
\cite{br}
\begin{equation} 
n^\mathit{exp}_c \; = \; 1.15 \pm 0.05 \label{nc}
\end{equation}
for the average number of charm (anti-)quarks per B$^+$/B$^0$-decay.
Complementary information on inclusive B decays can be obtained from
the semileptonic branching ratio. The CLEO and ARGUS groups
\cite{br,ex} have measured
\begin{eqnarray} 
B^\mathit{exp}_{SL} &=& 10.23 \pm 0.39\% . \label{bsl}
\end{eqnarray} 
The increasing experimental precision achieved in the current decade
has been paralleled by a substantial progress in the theoretical
understanding of the inclusive decay rates entering $B_{SL}$ and
$n_c$. Here the calculational key is the heavy quark expansion (HQE)
\cite{hqe1,hqe2} of the decay rate in question: The leading term of
the HQE reproduces the decay rate of a b-quark in the QCD-corrected
parton model. The first non-perturbative corrections are suppressed by
a factor of $(\Lambda_{QCD}/m_b)^2$ and affect the rates by at most a
few percent. Theoretically spectator effects of order $16 \pi^2
(\Lambda_{QCD}/m_b)^3$ \cite{spec,ns} could be larger \cite{ns}, but
for the decay rates of B$^\pm$ and B$^0$ entering \eq{bsl} and \eq{nc}
they are experimentally known to be at the percent level as well
\cite{kroll}.  The apparent smallness of these non-perturbative terms
has shifted the focus towards the perturbative corrections to the free
quark decay. The calculation of such short distance effects starts
from an effective hamiltonian, whose generic form reads
\begin{eqnarray} 
H &=& \frac{G_F}{\sqrt{2}} \lt[ V_{\mathit{CKM}} \sum_{j=1}^2 C_j Q_j -
       V_{\mathit{CKM}}^\prime \lt( \sum_{k=3}^6 C_j Q_j + C_8 Q_8 \rt) \rt]. 
\end{eqnarray} 
Here $G_F$ is the Fermi constant and $V_{\mathit{CKM}}$ and 
$V_{\mathit{CKM}}^\prime $ are products  of elements of the 
Cabibbo-Kobayashi-Maskawa (CKM) matrix. The Wilson coefficients
$C_j$ encode the physics connected to the weak scale and play the 
r\^{o}le of effective coupling constants of the local interactions 
described by the operators $Q_j$. Their precise form depends on the
flavour structure of the decay and will be given below in \eq{basis}. 

Decays with three different flavours in the final state such as
$b\rightarrow c \ov{u} d$ can only proceed through the current-current
operators $Q_1$ and $Q_2$. $\Gamma (b \rightarrow c \ov{u}
d)$ has been calculated to order \as, which is the next-to-leading
order (NLO), in \cite{bbbg1}. The generic Feynman diagram for these
corrections are depicted in \fig{fig:cc}.  In \cite{bbfg} the 
same diagrams have been calculated for $\Gamma ( b\rightarrow c \ov{c}
s )+\Gamma ( b\rightarrow c \ov{c} d)$. The latter decays and the 
charmless nonleptonic decays, however, also involve penguin effects.   
Diagrams with insertions of the penguin operators  $Q_{3-6}$ have been
taken into account only in the leading order (LO) \cite{bbbg2,bbfg,ap}, 
because their coefficients $C_{3-6}$ are much smaller than $C_{1,2}$
(cf.\ \tab{tab:wc}). The results for $B_{SL}$ and $n_c$ read
\begin{eqnarray}
B_{SL} = \lt( 11.7 \pm 1.4 \pm 1.0 \rt) \% , \qquad &&   
n_c = 1.34 \mp 0.06 .          \label{puzz}     
\end{eqnarray}
Here the result for $B_{SL}$ has been obtained in \cite{bbbg2,n} with
the analytical input from \cite{bbbg1,bbbg2,bbfg}. The second error
bar has been added to account for the spectator effects estimated in
\cite{ns}. Apparently there is no spectacular discrepancy between
\eq{bsl} and \eq{puzz}. The result for $n_c$ in \eq{puzz} does not involve
the calculation of $\Gamma ( b\rightarrow c \ov{c} d)+\Gamma (
b\rightarrow c \ov{c} s)$, but instead uses the experimental
information on $B_{SL}$ in \eq{bsl} as proposed in \cite{bdy,disy}.
We discuss this in more detail in sect.~\ref{sect:p}. 

The discrepancy between \eq{nc} and \eq{puzz} constitutes the
``missing charm puzzle''. The search for a theoretical explanation has
recently focused on new positive contributions to the yet unmeasured
charmless decay modes entering \eq{puzz}. Indeed, in a recent analysis
\cite{disy} $\Gamma (b\rightarrow \textit{no charm})$ has been
estimated indirectly in two different ways: First the experimental
information on final states with hadrons containing a $c$ quark has
been used and second data on decay products involving a $\ov{c}$ quark
have been analyzed. For the CLEO data both methods consistently
indicate an enhancement of $\Gamma (b\rightarrow \textit{no charm})$
by roughly a factor of 14 compared to the theoretical prediction in 
\cite{ap}.

Next we briefly discuss the LEP results for $B_{SL}$ and $n_c$
\cite{lep}.  The LEP Z-peak experiments encounter a mixture of
b-flavoured hadrons. In order to allow for a comparison with 
\eq{bsl} one must correct the LEP result 
$B^\mathit{Z,exp}_{SL} = 10.95 \pm 0.42 \% $ \cite{br}
for the different lifetimes \cite{kroll} of the hadrons in the mixture 
\cite{n2}:
\begin{eqnarray}
B^\mathit{Z,corr,exp}_{SL} &=& 11.13 \pm 0.42\% , \qquad \qquad \qquad 
n_c^\mathit{Z,exp} \; = \; 1.22 \pm 0.08 .\label{lepz}
\end{eqnarray}
These data are consistent with the theory (cf.~\eq{ncth} below), but
the analysis in \cite{disy} has found evidence for an enhanced $\Gamma (b\rightarrow \textit{no charm}) $
also from the LEP data. Further the two methods used in \cite{disy}
have given less consistent results for the LEP data than for the CLEO
data. In addition the LEP measurements involve the $\Lambda_b$ baryon,
whose lifetime is either not properly understood theoretically or
incorrectly measured. (If the latter is the case,
$B^\mathit{Z,corr,exp}_{SL}$ in \eq{lepz} must be replaced by the
uncorrected $B^\mathit{Z,exp}_{SL}$, which reduces the $2\sigma$
discrepancy between \eq{bsl} and \eq{lepz}.) Hence in our analysis we
will mainly use \eq{bsl} and \eq{nc}.

Now two possible sources of an enhanced $\Gamma (b\rightarrow
\textit{no charm}) $ are currently discussed: The authors of \cite{k}
stress the possibility that the Wilson coefficient $C_8$ of the
chromomagnetic dipole operator $Q_8$ is enhanced by new physics
contributions. On the other hand in \cite{disy,ps} an explanation
within QCD dynamics is suggested: An originally produced
$(c,\ov{c})$-pair can annihilate and thereby lead to a charmless final
state.

The calculation of matrix elements involving $Q_{3-6}$ and $Q_{8}$
does not exhaust all possible penguin effects.  In this paper we
calculate the contributions of penguin diagrams with insertions of the
current-current operator $Q_2$ to the decay rates into charmless final
states (see \fig{fig:peng}). Such a penguin diagram with a
$(c,\ov{c})$-pair in the intermediate state involves the large
coefficient $C_2$ and the CKM factor $V_{cb} \gg |V_{ub}|$. It is
precisely the short distance analogue of the mechanism proposed in
\cite{disy,ps} and surprisingly has not been considered in the
perturbative calculations \cite{bbfg,bbbg2,ap,acmp} of the decay rates
entering $B_{SL}$ and $n_c$. 

Further we calculate the diagrams involving the interference of the
tree diagram with $Q_8$ in \fig{fig:8} with $Q_{1-6}$, which belongs to
the order $\as$ as well.  The consideration of these diagrams is
mandatory, if one wants to estimate the effect of an enhanced
coefficient $C_8$ on $\Gamma (b\rightarrow \textit{no charm})$
proposed in \cite{k}.

The paper is organized as follows: In the following section we set up
our notations and collect results from earlier work.  The calculation
of the penguin diagram contributions and the $Q_{1-6}$--$Q_8$
interference terms is presented in sect.~\ref{sect:c}.  The
phenomenologically interested reader is refered to
sect.~\ref{sect:num}, in which we discuss our numerical results.  In
sect.~\ref{sect:num} we also comment on the mechanism proposed in
\cite{ps}.  Further a potential enhancement of $C_8$ is analyzed by a
model independent fit of $C_8$ to the experimental data. Finally we
conclude.

\begin{nfigure}{tb}
\begin{minipage}[t]{0.3\textwidth}
\centerline{\epsfxsize=0.6\textwidth \epsffile{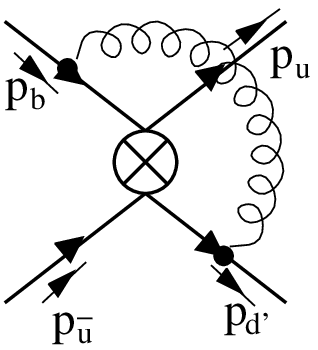}}
\caption{Example of a current-current diagram. The cross 
denotes the inserted operator. $d^\prime$ equals $d$ or $s$.}
\label{fig:cc}
\end{minipage}\hspace{2ex}
\begin{minipage}[t]{0.3\textwidth}
\centerline{\epsfysize=0.6\textwidth \epsffile{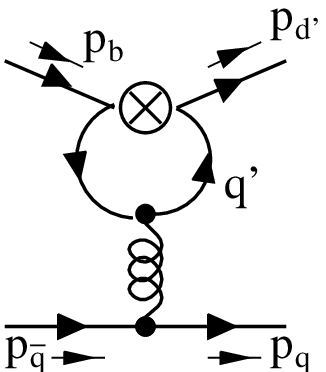}}
\caption{Penguin diagram involving $Q_2$. The internal quark 
$q^\prime$ can be $u$ or $c$. The corresponding diagram with $Q_1$
vanishes.}
\label{fig:peng}
\end{minipage}\hspace{2ex}
\begin{minipage}[t]{0.3\textwidth}
\centerline{\epsfysize=0.6\textwidth \epsffile{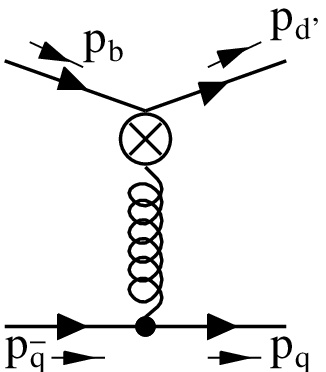}}
\caption{Tree diagram contribution of $Q_8$ to 
  $\Gamma (b\rightarrow q\ov{q} d^\prime )$. }
\label{fig:8}
\end{minipage}
\end{nfigure}

\section{Preliminaries}\label{sect:p}
\subsection{$\mathbf{B_{SL}}$ and $\mathbf{n_c}$} 
For the theoretical description of the various decay rates it is
advantageous to normalize them to the well-understood semileptonic
decay rate \cite{bdy,disy}: 
\begin{eqnarray} 
\!\!\! r_{q\ell}  =  
   \frac{\Gamma \lt( \ov{B} \rightarrow X_q \ell \ov{\nu}_{\ell} \rt) }{
         \Gamma \lt( \ov{B} \rightarrow X_c e \ov{\nu}_e \rt) } 
, \qquad
\rqq{q_1}{q_2}{q_3} \!&=&\! 
        \frac{\Gamma \lt( \ov{B} \rightarrow X_{q_1\ov{q_2}q_3} \rt) }{
              \Gamma \lt( \ov{B} \rightarrow X_c e \ov{\nu}_e \rt) } 
, \qquad 
r_{qg}  =   \frac{\Gamma \lt( \ov{B} \rightarrow X_{qg} \rt) }{
              \Gamma \lt( \ov{B} \rightarrow X_c e \ov{\nu}_e \rt) }  .
\label{defr}
\end{eqnarray} 
This eliminates the factor of $m_b^5$ common to all decay rates.  For
the charmless decays we will further use
\begin{eqnarray} 
\rnc &=& \sum_{ \scriptstyle
                 q = d,s \atop  \scriptstyle q^\prime = u,d,s 
              } \lt[  
         r_{q^\prime\ov{q}^\prime q} + r_{qg}  \rt] + 
         2 r_{u e} . \label{defrnc}
\end{eqnarray}
The semileptonic branching ratio reads 
\begin{eqnarray} 
B_{SL} &=& \frac{1}{
 2 + r_{c\tau} + \rnc + \sum_{q = d,s } 
        \lt[ \rqq{c}{u}{q} + \rqq{c}{c}{q}  
             \rt]}  . \label{bslth}
\end{eqnarray} 
Small contributions such as $r_{u \tau}=0.004$ and radiative decay
modes have been omitted in \eq{defrnc}.  In \eq{bslth} also
$r_{u\ov{c}s}=0.05$ \cite{ap} has been neglected.
The numerical value for $B_{SL}$ in \eq{puzz} is the average of the
two results given in \cite{bbbg2} corresponding to two different
renormalization schemes (on-shell vs.\ $\ov{\rm MS}$ \cite{bbdm} quark
masses).  In the corresponding expression for $n_c$ the term
$\rqq{c}{c}{s}+\rqq{c}{c}{d}$, which suffers from sizeable theoretical
uncertainties, is eliminated in favour of $B_{SL}$ \cite{bdy,disy}:
\begin{eqnarray}  
n_c &=& 2 - \lt( 2+ r_{c\tau} + \rqq{c}{u}{d} +  \rqq{c}{u}{s} + 2 \rnc
              \rt) B_{SL} . 
\label{ncth}
\end{eqnarray}
Yet with \cite{bbbg1,flnn}
\begin{eqnarray}
\rqq{c}{u}{s}+\rqq{c}{u}{d} \; = \; 4.0 \pm 0.4, && \qquad
r_{c\tau} \; = \; 0.25    
\end{eqnarray}
and $B^\mathit{exp}_{SL}$ in \eq{bsl} one obtains 
\begin{eqnarray}
n_c &=& 1.36 \mp 0.04 \, - \,  0.205 \cdot \rnc  \label{ncnc}.
\end{eqnarray} 
Now the current-current type radiative corrections to $\rnc$
(cf.~\fig{fig:cc}) have been calculated in \cite{acmp,b,bbbg1} using
different renormalization schemes. The penguin operators $Q_{3-6}$
have been included within the LO in \cite{ap}.  With up-to-date values
for $B_{SL}$ and the CKM elements the calculation of \cite{ap} yields
$\rnc = 0.11 \pm 0.08$.  Inserting this result into \eq{ncnc} yields
the numerical value in \eq{puzz}. In contrast the indirect
experimental determination in \cite{disy} has found $\rnc = 1.6 \pm
0.4$. In order to reproduce the experimental value of $n_c$ in \eq{nc}
one needs $\rnc=1.0\pm 0.4$.

So far the penguin diagrams of \fig{fig:peng} have not been calculated
for all possible charmless B decay modes.  Yet for the pure penguin
induced decays of a $b$-quark into three down-type (anti)-quarks the
contribution of the diagram of \fig{fig:peng} has been obtained in
terms of a twofold integral representation in \cite{f}. Likewise the
effect of penguin diagrams on the analysis of CP asymmetries has been
studied in \cite{f,bf} and in \cite{cfms} penguin effects on exclusive
decays have been studied.

\subsection{Decay rates to order $\as$}
In order to describe decays of the type $b\rightarrow q\ov{q} d$ one
needs the following hamiltonian:
\begin{eqnarray}
H &=&  \frac{G_F}{\sqrt{2}} 
        \left\{ 
        \sum_{j=1}^2 C_j \left( \xi_c^* Q_j^c + 
                                \xi_u^* Q_j^u \right) 
        - \xi_t^*    
          \sum_{j \in \mathcal{P}} C_j Q_j 
        \right\} \; , \qquad \qquad
\xi_{q^\prime} \; = \; V_{q^\prime b}^* V_{q^\prime d}
\; .
\label{hd}
\end{eqnarray} 
Here $\xi_u + \xi_c + \xi_t =0$ due to the unitarity of the CKM
matrix and $\mathcal{P}= \{ 3,\ldots 6, 8 \}$. $H$ in \eq{hd} comprises
the following operator basis:\footnote{The overall sign of the matrix
  element $\langle Q_{8} \rangle $ depends on the chosen sign of the
  coupling $g$ in the covariant derivative in the QCD lagrangian.  The
  definition in $\eq{basis}$ complies with the result in \cite{cm,cfmrs}, if
  the covariant derivative is chosen as $D_\mu =
  \partial_\mu - i g T^a A^a_\mu$, so that the Feynman rule for the
  fermion-gluon vertex is $i g T^a$. By convention the notation $Q_7$ is  
  reserved for the magnetic $\ov{d}b\gamma$-operator, which we do not need 
  in our calculation.
}
\begin{eqnarray}
&&
\left.
\begin{array}{@{\hspace{-14em}}rcl}
Q_1^{q} &=& (\bar{d}q)_{V-A}  \cdot (\bar{q}b)_{V-A} \cdot \tw \no \\[1mm]
Q_2^{q} &=& (\bar{d}q)_{V-A}  \cdot (\bar{q}b)_{V-A} \cdot \1 \nn
\end{array}
\right\} \quad \textrm{with $q=u$ or $q=c$,}
\nn
Q_3^{q} \; = \;  (\bar{d}b)_{V-A}  \cdot (\bar{q}q)_{V-A}
  \cdot \1   ,
 && \qquad Q_4^{q} \; = \; (\bar{d}b)_{V-A}  \cdot (\bar{q}q)_{V-A}
  \cdot \tw 
\nn
Q_5^{q} \; = \; (\bar{d}b)_{V-A}  \cdot (\bar{q}q)_{V+A}
  \cdot \1 ,
 && \qquad Q_6^{q} \; = \; (\bar{d}b)_{V-A}  \cdot (\bar{q}q)_{V+A}
  \cdot \tw 
\nn 
Q_j \; = \; \hspace{-2ex} \sum_{q=u,d,s,c,b} \hspace{-2ex}
Q_j^{q}  \hspace{6em}
 && \qquad \mbox{for $3\leq j\leq 6$,}
 \nn 
 Q_{8} \; =\;  - \frac{g}{8 \pi^2} \, \bar{d} \sigma^{\mu \nu} 
          \left[ m_d L + m_b R  \right] T^a b \cdot G^a_{\mu \nu} 
 \; . \hspace{-6em}
\label{basis}
\end{eqnarray}
The colour singlet and non-singlet structure are indicated by \1\ and
\tw\ and $V\pm A$ is the Dirac structure, i.e.\ 
\begin{eqnarray}
\lt( \bar{d}q \rt)_{V-A}  \cdot \lt( \bar{q}b \rt)_{V-A} \cdot \tw
&=& \bar{d}_{\alpha} \g_\mu \lt( 1-\g_5 \rt) q_{\beta} \cdot 
    \bar{q}_{\beta} \g^\mu \lt( 1-\g_5 \rt) b_{\alpha}. \no
\end{eqnarray}
Next it is useful to expand the renormalized matrix elements in
\as\ and to separate the result from current-current diagrams (see
\fig{fig:cc}) and penguin diagrams (see \fig{fig:peng}):
\begin{eqnarray}
\bra{q\ov{q} d} Q_j^{q^\prime} \ket{b} &=& 
   \langle \, Q_j^{q^\prime} \, \rangle^{(0)}+ 
    \frac{\as}{4 \pi} \left[ 
     \langle \, Q_j^{q^\prime} \, \rangle^{(1)}_\mathit{cc} + 
     \langle \, Q_j^{q^\prime} \, \rangle^{(1)}_\mathit{peng} \right] +
    O(\as^2), \qquad j=1,2, 
\label{qexp}  \\
\langle \, Q_j^{q^\prime} \, \rangle^{(1)}_\mathit{peng}
  &=& \sum_{k \in \mathcal{P}} r^{q^\prime}_{jk} 
      \lt( p^2, m_{q^\prime},\mu \rt)  \langle Q_k
  \rangle^{(0)},  \qquad \qquad \qquad p=p_b-p_d \;. \label{r}
\end{eqnarray}
Of course $\langle \, Q_j^{q^\prime} \, \rangle^{(0)}$ and $\langle \,
Q_j^{q^\prime} \, \rangle^{(1)}_\mathit{cc}$ are non-zero only for
$q=q^\prime=u$, recall that we do not consider $q=c$ in this work.  In
\eq{r} we have expressed the result of the penguin diagram in terms of
the tree-level matrix elements. There $\mu$ is the renormalization
scale. For the momentum flow cf.~\fig{fig:peng}.

The quark decay rate is related to the matrix element of $H$ via
\begin{eqnarray}
\Gamma \lt( b\rightarrow q \ov{q} d \rt) 
&=& \frac{1}{2 m_b} \! 
           \int \! \! 
           \frac{d^3 \vec{p}_q d^3 \vec{p}_{\bar{q}} 
        d^3 \vec{p}_d}{(2\pi)^5 8 |E_q E_{\bar{q}} E_d| }  
     \delta^{(4)} \left( p_b + p_{\bar{q}} - p_q -p_d \right)
     \ov{ \bra{q \ov{q}d} H \ket{b} \bra{q \ov{q}d} H \ket{b}^{*} } .\no
\end{eqnarray}
The bar over $\bra{q \ov{q}d} H \ket{b} \bra{q \ov{q}d} H \ket{b}^{*}$
denotes the average over initial state polarizations and the sum over
final state polarizations.  Next we expand the decay rate to order \as
:
\begin{eqnarray}
\Gamma \lt( b\rightarrow q \ov{q} d \rt) &=& 
   \Gamma^{(0)}+ 
     \frac{\as (\mu)}{4 \pi} 
            \lt[ \Delta \Gamma_\mathit{cc}  + 
                 \Delta \Gamma_\mathit{peng} +
                 \Delta \Gamma_8 + \Delta \Gamma_\mathit{W} \rt] + O(\as^2)
. \label{gexp}
\end{eqnarray}
For $b\rightarrow q \ov{q} s$ decays one simply substitutes $d$ by $s$
in (\ref{basis}-\ref{gexp}).  Now the first two terms of the NLO
correction in \eq{gexp} describe the effect of current-current and
penguin diagrams involving $Q_1$ or $Q_2$. $\Delta \Gamma_8$ likewise
contains the matrix elements of $Q_8$.  The remaining part $\Delta
\Gamma_W$ of the NLO contribution is made of the corrections to the
Wilson coefficients \cite{acmp,bjlw} multiplying the tree-level
amplitudes in $\Gamma^{(0)}$. We write
\begin{eqnarray}
C_j (\mu) &=& C_j^{(0)} (\mu) + \frac{\as (\mu)}{4 \pi} 
              \Delta C_j (\mu) , \qquad\qquad j=1,\ldots 6. \label{wc}
\end{eqnarray} 
Here $\Delta C_j$ is the NLO correction to the Wilson coefficient.
$\Delta C_j$ depends on the renormalization scheme chosen.  This
scheme dependence cancels with a corresponding one in the results of
the loop diagrams contained in $\Delta \Gamma_{cc}$ and $\Delta
\Gamma_{peng}$.  For example the scheme dependence of $\Delta C_{1,2}$
cancels in combination with the current-current type corrections to
$Q_1$ and $Q_2$ of \fig{fig:cc}. Since we do not include the unknown
radiative corrections to the penguin operators $Q_{3-6}$ in \eq{gexp},
we must likewise leave out terms in $\Delta C_j$ related to the NLO
penguin-penguin mixing in order to render $\Gamma$ in \eq{gexp} scheme
independent.  We ban these technical details into the appendix.  The
values of the Wilson coefficients needed for the numerical evaluation
of the various decay rates are listed in \tab{tab:wc}.

\begin{ntable}{tb}
\begin{tabular}{l||r|r|r|r|r|r|r}
j & 1&2&3&4&5&6&8 \\[0.4ex]\hline
$C_j^{(0)} (\mu=m_b)$ & -0.2493 & 1.1077 & 0.0111 & -0.0256 & 0.0075 &
                        -0.0315 & -0.1495 \\[0.2ex]
$C_j^{NDR} (\mu=m_b)$ &   -0.1739 &  1.0731  & 0.0113 &  -0.0326 & 0.0110 
                      & -0.0384 & \\[0.4ex]\hline 
$C_j^{(0)} (\mu=m_b/2)$ & -0.3611 & 1.1694 &  0.0170 &  -0.0359 &  0.0100 &
                        -0.0484 & -0.1663 \\[0.2ex] 
$C_j^{NDR} (\mu=m_b/2)$ &  -0.2720 &  1.1246 &   0.0174 &  -0.0461 &
                             0.0149 &    -0.0587 &  \\[0.4ex]\hline
$C_j^{(0)} ( \mu=2 \,m_b)$ &  -0.1669 &  1.0671  &  0.0071  &  -0.0176 & 0.0054
                        & -0.0202 &  -0.1355 \\[0.2ex]
$C_j^{NDR} (\mu=2 \,m_b)$ &  -0.1001 & 1.0389 &  0.0073 &  -0.0227 &  0.0079
                        &  -0.0251 & \\[0.4ex]\hline
$\Delta \ov{C}_j (\mu =m_b )$ & 2.719 & -1.744 & 0.380 & -0.1050  
                              & -0.223 & 0.384 & 
\end{tabular}

\caption{Wilson coefficients used in our analysis. $C_j^{(0)}$ is the 
  LO expression, $C_j^{NDR}$ is the NLO coefficient in the NDR scheme.
  In $C_j^{NDR}$ above the NLO corrections to penguin-penguin mixing
  have been omitted in order to render $\Gamma$ in \eq{gaall} scheme
  independent as described in the text. For $\mu=m_b$ this affects
  $C_3$ and $C_5$ by 12 \% and 25 \%, but is negligible for the other
  coefficients. $C_j^{(0)}$ and $C_j^ {NDR}$ are needed for the
  numerical evaluation of the decay rate in \eq{gaall}.  $\Delta
  \ov{C}_j $ in the last line is defined in \eq{wca}.  The top and
  bottom mass are chosen as $m_t=m_t^{\ov{\mathrm{MS}}}(m_t)=168$ GeV
  and $m_b=4.8$ GeV. Further $\as (M_Z) = 0.118$ \cite{beth}, which
  corresponds to $\as (4.8 \mathrm{ GeV}) = 0.216$.  In the table
  $C_8^{(0)}=C_8^{(0),eff}$ is the scheme independent coefficient
  mentioned in the appendix.  }\label{tab:wc}
\end{ntable}

In the LO the decays $b\rightarrow s \ov{s} s$, $b\rightarrow s \ov{s} d$, 
$b\rightarrow d \ov{d} s$ and $b\rightarrow d \ov{d} d$ can only
proceed via $Q_{3-6}$ and $Q_8$, while  $b\rightarrow u \ov{u} d$ and 
$b\rightarrow u \ov{u} s$ also receive contributions from $Q_1$ and
$Q_2$. We combine both cases in 
\begin{eqnarray} 
  \Gamma^{(0)} &=& \frac{G_F^2 m_b^5}{64 \pi^3} \lt[ \,t
  \sum_{i,j=1}^2 \lt| \xi_u \rt|^2 C_i^{(0)} C_j^{(0)} b_{ij} + 
  \sum_{i,j=3}^6 \lt| \xi_t \rt|^2 C_i^{(0)} C_j^{(0)} b_{ij} \rt. \nn
&& \phantom{  \frac{G_F^2 m_b^5}{192 \pi^3} \big[ } \;
 \lt. - 2 \, t \sum_{ \scriptstyle i = 1,2 \atop 
           \scriptstyle j = 3, \ldots 6} C_i^{(0)} C_j^{(0)} \,
  \real \lt( \xi_u^* \xi_t \rt) \, b_{ij} \rt] \label{gtree}
\end{eqnarray}
with $t=1$ for $q=u$ and $t=0$ for $q=d,s$. The $b_{ij}$'s read
\begin{eqnarray}
b_{ij} &=& \frac{16 \pi^3}{m_b^6 } \int 
  \frac{d^3 \vec{p}_q d^3 \vec{p}_{\bar{q}} 
        d^3 \vec{p}_d}{(2\pi)^5 8 |E_q E_{\bar{q}} E_d| }  
   \delta^{(4)} \left( p_b + p_{\bar{q}} - p_q -p_d \right)
   \ov{ \langle Q_{i} \rangle^{(0)} \langle Q_{j} \rangle^{(0)\,*} } 
\; =\; b_{ji} 
\label{defbijc}
\end{eqnarray}
with $Q_{1,2}=Q_{1,2}^u$ here.  Setting the final state quark masses
to zero one finds
\begin{eqnarray} 
b_{ij} &=& \left\{
\begin{array}{ll}
1+r/3   & \mbox{ for } i,j \leq 4,\mbox{ and } i+j \mbox{ even } \\ 
1/3 +r & \mbox{ for } i,j \leq 4,\mbox{ and } i+j \mbox{ odd } 
\end{array} \right\}, 
\qquad \qquad 
\begin{array}{l}
b_{55} \;=\; b_{66} \; = \; 1 \; , \\   
b_{56} \;=\; b_{65} \; = \; 1/3 \; . 
\end{array} 
\label{bij}
\end{eqnarray}
Here $r=1$ for the decays $b\rightarrow d \ov{d} d$ and $b\rightarrow
s \ov{s} s$, in which the final state contains two identical
particles, and $r=0$ otherwise.  The remaining $b_{ij}$'s are zero.
Clearly for the $q\neq u$ the $b_{ij}$'s as defined in \eq{defbijc}
vanish, if $i \leq2$ or $j\leq2$. Yet in the formulae for the decay
rate we prefer to stress this fact by keeping the parameter $t$, which
switches the current-current effects off in the penguin induced decays.
Our results in \eq{bij} agree with the zero mass limit of
\cite{bbbg2,bbfg}.  The $b_{ij}$'s of \eq{defbijc} are visualized in
\fig{fig:inc}.

\begin{nfigure}{tb}
\centerline{\epsfxsize=0.9\textwidth \epsffile{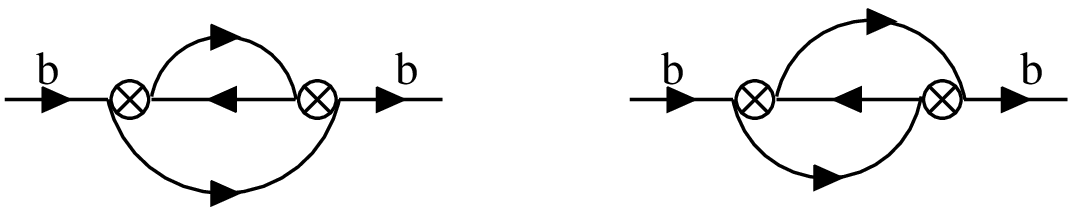}}
\caption{The two diagrams contributing to $\Gamma^{(0)}$ in
  \eq{gtree}.  The crosses represent any of $Q_{1-6}$. }
\label{fig:inc}
\end{nfigure}

$\Delta \Gamma_{W}$ simply reads
\begin{eqnarray} 
\Delta \Gamma_{W} \hspace{-3pt} &=& \hspace{-3pt}
   \frac{G_F^2 m_b^5}{64 \pi^3}\, 2 \lt[
  t \sum_{i,j=1}^2 \lt| \xi_u \rt|^2 
 \lt[ C_i^{(0)} \Delta C_j  \rt]
 b_{ij}  + 
  \sum_{i,j=3}^6 \lt| \xi_t \rt|^2 
  \lt[ C_i^{(0)} \Delta C_j  \rt]
 b_{ij} \rt. \nn
&& \phantom{ 2\,  \frac{G_F^2 m_b^5}{192 \pi^3} \big[ } \;
 \lt. - \, t \sum_{ \scriptstyle i = 1,2 \atop \scriptstyle j = 3, \ldots 6} 
  \lt[ C_i^{(0)} \Delta C_j + \Delta C_i C_j^{(0)}  \rt]
  \real \lt( \xi_u^* \xi_t \rt) \, b_{ij}  \rt] \label{gw}.
\end{eqnarray}

The current-current type corrections proportional to 
$C_{1,2}^{(0)} \cdot C_{1,2}^{(0)}$ are \cite{acmp,b,bbbg1} 
\begin{eqnarray}
\Delta \Gamma_{cc} &=& t \,  \frac{G_F^2 m_b^5}{64 \pi^3} 
   \, 2 \, \left| \xi_u \right|^2 \sum_{i,j=1}^2 
   C_i^{(0)} C_j^{(0)} h_{ij} 
\label{cc}
\end{eqnarray}
with $t$ defined after \eq{gtree}. In the NDR scheme with the standard
definition of the evanescent operators \cite{hn} the diagrams of
\fig{fig:cc}, the bremsstrahlung diagrams and the subsequent phase
space integrations yield \cite{bbbg1}:
\begin{eqnarray}
\!\!
h^\mathit{NDR}_{11} = h^\mathit{NDR}_{22} = 
            \frac{31}{3} -\frac{4}{3} \pi^2,  && \quad 
h^\mathit{NDR}_{12} \lt( \frac{\mu}{m_b} \rt) = 
h^\mathit{NDR}_{21} \lt( \frac{\mu}{m_b} \rt) = 
             \frac{8}{3} \ln \frac{m_b^2}{\mu^2} - \frac{17}{3} -
               \frac{4}{9} \pi^2  .\label{defh}
\end{eqnarray}
The inclusion of $\Delta \Gamma_{cc}$ is necessary to fix the
definiton of the b-quark mass entering $\Gamma^{(0)}$ in \eq{gtree}.
The values quoted in \eq{defh} correspond to the use of the (one-loop)
pole quark mass in \eq{gtree}.  

In the same way we write 
\begin{eqnarray} 
\Delta \Gamma_\mathit{peng} &=&   \frac{G_F^2 m_b^5}{64 \pi^3} 
   \, 2 \,  \real \lt[ \, t  
    \sum_{i,j=1,2} C_i^{(0)} C_j^{(0)} \,    \xi_u 
    \lt[ \xi_c^* g_{ij} (x_c)  + \xi_u^* g_{ij} (0) \rt]
     \rt.  \nn
&& \phantom{   \frac{G_F^2 m_b^5}{64 \pi^3} 
   \, 2 \, \real  }  \lt. \;\; 
- \! \sum_{ \scriptstyle i = 1,2 \atop \scriptstyle 
          j = 3, \ldots 6} \! C_i^{(0)}  C_j^{(0)} 
    \xi_t 
    \lt[ \xi_c^* g_{ij}(x_c)  + \xi_u^* g_{ij} (0) \rt] \rt] . \label{peng}
\end{eqnarray} 
Here $g_{ij}$ is visualized in \fig{fig:pinc} and defined by 
\begin{eqnarray}
\! 
g_{ij} \lt( x_{q^\prime} , \frac{\mu}{m_b} \rt) 
 \!\!  &=& \!\! \frac{16 \pi^3}{m_b^6} 
      \int \frac{d^3 \vec{p}_q d^3 \vec{p}_{\bar{q}} 
        d^3 \vec{p}_d}{(2\pi)^5 8 |E_q E_{\bar{q}} E_d| }  
   \delta^{(4)} \left( p_b + p_{\bar{q}} - p_q -p_d \right) \cdot 
   \ov{ \langle \, Q_i^{q^\prime} \, \rangle^{(1)}_\mathit{peng} 
        \langle Q_{j} \rangle^{(0)\,*} }  \label{defgij}
\end{eqnarray}
with $x_{q^\prime}=m_{q^\prime}/m_b$. We do not display the
$\mu$-dependence of the $C_j$'s, $h_{ij}$'s and $g_{ij}$'s in formulae
for the decay rate such as \eq{cc} or \eq{peng} to simplify the
notation.  Now $\Delta \Gamma_\mathit{peng}$ in \eq{peng} is more
complicated than $\Delta \Gamma_\mathit{cc}$ in \eq{cc} for two
reasons: First interference terms of different CKM structures appear
and second the internal quark in the penguin graph of \fig{fig:peng}
can be $q^\prime=c$ or $q^\prime=u$. Further the charm quark mass
enters $g_{ij}$ with $x_c=m_c/m_b$.

\begin{nfigure}{tb}
\centerline{\epsfxsize=0.9\textwidth \epsffile{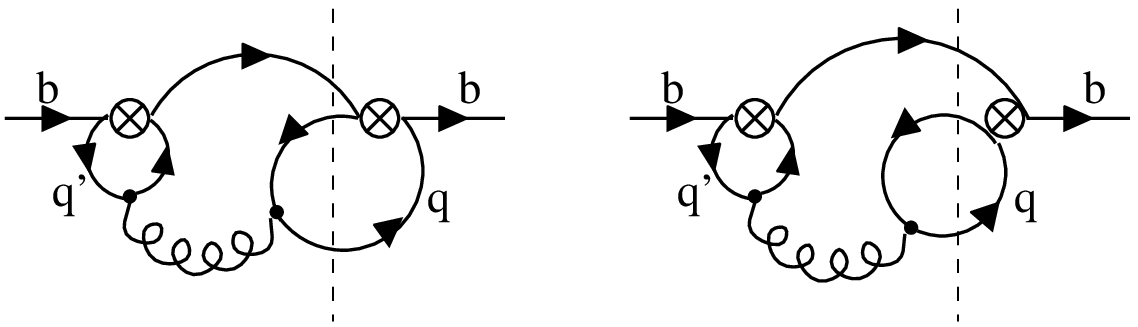}}
\caption{The diagrams contributing to $\Delta \Gamma_\mathit{peng}$
  in \eq{peng}.  The left cross denotes $Q_1^{q^\prime}$ or
  $Q_2^{q^\prime}$ with $q^\prime=u,c$ and the right cross represents
  any of $Q_{1-6}$. The dashed line indicates the final state with $q$
  being $u,d$ or $s$.  }
\label{fig:pinc}
\end{nfigure}

Finally $\Delta \Gamma_8$ in \eq{gexp} is given by 
\begin{eqnarray}
\Delta \Gamma_8 &=& 
 \frac{G_F^2 m_b^5}{64 \pi^3} 
    \, 2 \,  \real \! \lt[ - t \, \xi_u^* \xi_t C^{(0)}_8 \sum_{j=1}^2 
    C^{(0)}_j b_{j8} + \lt| \xi_t \rt|^2 C^{(0)}_8 
     \sum_{j=3}^6 C^{(0)}_j b_{j8} \rt] . \label{ga8}
\end{eqnarray}
Here the tree-level diagrams with $Q_8$ already contribute to 
order \as.

The phase space integrations are contained in the coefficients
$b_{j8}$ in \eq{ga8}. They are depicted in \fig{fig:8inc} and are
defined as 
\begin{eqnarray} 
b_{j8} &=& \frac{16 \pi^3}{m_b^6 } \frac{4\pi}{\as}
 \int \frac{d^3 \vec{p}_q d^3 \vec{p}_{\bar{q}} 
        d^3 \vec{p}_d}{(2\pi)^5 8 |E_q E_{\bar{q}} E_d| }  
   \delta^{(4)} \left( p_b + p_{\bar{q}} - p_q -p_d \right)
   \ov{ \langle Q_{j} \rangle^{(0)} \langle Q_{8} \rangle^{(0)\,*} } 
\; =\; b_{8j} .
\label{defbj8}
\end{eqnarray} 

\begin{nfigure}{tb}
\centerline{\epsfxsize=0.9\textwidth \epsffile{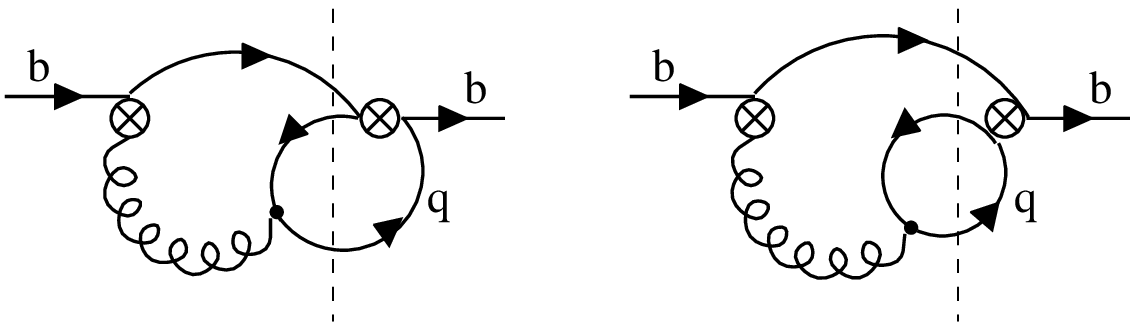}}
\caption{The diagrams contributing to $\Delta \Gamma_8$
  in \eq{ga8}.
  The left cross denotes $Q_8$ and the right cross  
  represents any of $Q_{1-6}$.}
\label{fig:8inc}
\end{nfigure}

It is instructive to insert the above expressions for $\Delta
\Gamma_{cc}$, $\Delta \Gamma_{peng}$ and $\Delta \Gamma_{8}$ into
\eq{gexp}. The decay rate then reads
\begin{eqnarray}
\Gamma &=&  \frac{G_F^2 m_b^5}{64 \pi^3} 
    \,   \real \lt\{ t \,  \sum_{i,j=1}^2  C_i C_j 
     \lt[ \lt| \xi_u \rt|^2 b_{ij} +  
     \frac{\alpha_s \lt( \mu \rt)}{4 \pi} \lt| \xi_u \rt|^2 
         \, 2\,  \lt[ h_{ij} + g_{ij} (0)- g_{ij} (x_c) \rt] \rt.\rt. \nn 
&& \phantom{ \frac{G_F^2 m_b^5}{64 \pi^3} 
    \, 2 \,  \real \{ t \,  \sum_{i,j=1}^2  C_i C_j 
    [ \lt| \xi_u \rt|^2 b_{ij} } \;
\lt.       -  \frac{\alpha_s \lt( \mu \rt)}{4 \pi} \xi_u \xi_t^* 
          \, 2 \,g_{ij} ( x_c ) \rt]  \nn
&& \phantom{ \frac{G_F^2 m_b^5}{64 \pi^3} } 
   - 2 \! \sum_{ \scriptstyle i = 1,2 \atop \scriptstyle j = 3, \ldots
    6} 
       \! C_i  C_j 
      \lt[ t\, \xi_u^* \xi_t \, b_{ij} +  \frac{\alpha_s \lt( \mu \rt)}{4 \pi}
      \xi_u^* \xi_t \lt[ g_{ij} (0) - g_{ij} (x_c) \rt]
   - \frac{\alpha_s \lt( \mu \rt)}{4 \pi} \lt| \xi_t \rt|^2 g_{ij} (x_c) 
      \rt] \nn 
&& \phantom{ \frac{G_F^2 m_b^5}{64 \pi^3} } \;
   + \sum_{i,j=3}^6 C_i C_j \, \lt| \xi_t \rt|^2 \, b_{ij} \nn
&& \phantom{ \frac{G_F^2 m_b^5}{64 \pi^3} } \;
\lt.   +  \frac{\alpha_s \lt( \mu \rt)}{4 \pi} C_8
     \lt[ -t \, \xi_u^* \xi_t \sum_{j=1}^2 C_j \, 2 \,b_{j8} 
        + \lt| \xi_t \rt|^2 \sum_{j=3}^6 C_j \, 2\, b_{j8} \rt] \rt\}
. 
\label{gaall}
\end{eqnarray}
Here the $C_j$'s are the Wilson coefficients of \eq{wc} including NLO
corrections.  $C_j^{(0)}$ rather than $C_j^{NDR}$ should be used in
the terms of order $\as$ in \eq{gaall} for consistency.  The unitarity
of the CKM matrix has been used to eliminate $\xi_c$ from \eq{gaall}.

From \eq{peng} or \eq{gaall} one notices that penguin induced decays
with $t=0$ receive radiative corrections proportional to the large
coefficient $C_2$, which does not enter the tree-level decay rate in
\eq{gtree}.  Further $\Delta \Gamma_{peng}$ depends on the CKM phase
$\delta$, because $\xi_u$, $\xi_t$ and the loop functions $g_{ij}$ are
complex.  Decay rates and CP asymmetries for these penguin induced
decays have been derived in \cite{f} in terms of a two-fold integral
representation taking into account the interference of the penguin
diagram involving $Q_2$ with $Q_{1-6}$.  In sect.~\ref{sect:c} we
derive analytical results for the decay rates and also include $\Delta
\Gamma_{8}$.

In the decays $b\rightarrow u \ov{u} d$ and $b\rightarrow u \ov{u} s$
the main focus is on the first sum in $\Delta \Gamma_{peng}$ in
\eq{peng} containing products of two of the large coefficients $C_1$
and $C_2$. We also keep the second sum in \eq{peng}, but remark here
that these terms are not the full set of one-loop radiative
corrections involving one large coefficient $C_{1,2}$ and one small
coefficient $C_{3-6}$: In addition to the radiative corrections
calculated in this paper there are also current-current diagrams (see
\fig{fig:cc}) and penguin diagrams (see \fig{fig:peng}) with penguin
operators $Q_{3-6}$. In decays with a $(u,\ov{u})$-pair in the final
state these matrix elements interfere with those of $Q_1^u$ and
$Q_2^u$ and therefore also yield a term proportional to $C_{1,2}\cdot
C_{3-6}$. 

The terms proportional to $C_j \cdot C_8$ comprised in $\Delta
\Gamma_8$ are interesting in order to confirm or falsify the mechanism
proposed in \cite{k}. If new physics indeed dominates $C_8$, then
$\Gamma (b\rightarrow s+g)\propto |C_8|^2$ considered in \cite{k} is
enhanced. Yet $\Delta \Gamma_{8}$ is linear in $C_8$, so that the sign
of the new physics contribution may determine whether $\Gamma
(b\rightarrow \textit{no charm})$ is enhanced or diminished.  A
calculation similar to ours has been partly done in \cite{sew}. Yet in
\cite{sew} some questionable approximations have been made: For
example the operator mixing has been neglected and the top quark has
not been integrated out but instead formally treated as a light
particle.  In some cases the results of \cite{sew} differ
substantially from those in \cite{f}.

We close this section with the formula relating $\Gamma$ in \eq{gaall}
to the $\rqq{q_1}{q_2}{q_3}$'s defined in \eq{defr}. To this end 
we need the semileptonic decay rate to order $\as$ \cite{nir}:
\begin{eqnarray}
\Gamma \lt( b \rightarrow c e \ov{\nu}_e \rt) &=&  
       \frac{G_F^2 m_b^5}{192 \pi^3} \lt| V_{cb} \rt|^2 
        f_1 \lt( x_c^2 \rt) \lt[ 1 + \frac{\as (\mu) }{2 \pi} \,
       h_{SL} (x_c) + O \lt( \as^2 \rt) \rt]  . \no
\end{eqnarray}
The tree-level phase space function is 
\begin{eqnarray}
f_1 (a) &=& 1 - 8 a - 12 a^2 \ln a + 8 a^3 - a^4 . \no 
\end{eqnarray}
The analytic expression for $h_{SL} (x_c)$ can be found in
\cite{bbbg1,nir}. The approximation
\begin{eqnarray} 
h_{SL} \lt( x_c \rt) &=& -3.341 + 4.05  \lt( x_c-0.3 \rt) - 
                         4.3 \lt( x_c-0.3 \rt)^2 \no
\end{eqnarray}
holds to an accuracy of 1 permille in the range $0.2 \leq x_c \leq
0.4$. Here $x_c=m_c/m_b$ is the ratio of the one-loop pole masses.
We further include the hadronic corrections to the free quark
decay of order $1/m_b^2$ obtained from the HQE \cite{hqe1}. This
yields
\begin{eqnarray}
\!\!\!\!\!\!
\rqq{q}{q}{d} \hspace{-0.9ex} &=& \hspace{-0.9ex}
  \frac{192 \pi^3}{G_F^2 \, m_b^5\, \lt|V_{cb} \rt|^2 f_1\lt(x_c^2\rt)} 
     \lt\{ \Gamma + \Gamma^{(0)} \! 
           \lt[ 6 \lt( \frac{\lt(1-x_c^2\rt)^4}{f_1 \lt( x_c^2 \rt)}
                   -1 \rt) \frac{\lambda_2}{m_b^2} - 
                   \frac{\alpha_s ( \mu )}{2 \pi} h_{SL} \lt( x_c \rt)
           \rt] \! + \delta \Gamma  \rt\}\!\! . \label{rqq}
\end{eqnarray}
Here $\lambda_2=0.12 $ GeV$^2$ encodes the chromomagnetic interaction
of the b-quark with the light degrees of freedom. Further corrections
are contained in $ \delta \Gamma $. It is obtained from $\Gamma^{(0)}$
by substituting $b_{ij}$ with $\delta b_{ij}$ in the definition
\eq{gtree}. One finds from \cite{hqe1}:
\begin{eqnarray}
\delta b_{12} = \delta b_{14} = 
\delta b_{21} = \delta b_{23}   =  \delta b_{32} =  
\delta b_{34} =  \delta b_{41} =  
 \delta b_{43} \; = \; - 8 \, \frac{\lambda_2}{m_b^2} 
        & = & -0.042 \nn && \mbox{for $m_b=4.8$ GeV}, \no
\end{eqnarray}
while $\delta b_{56}=\delta b_{65}$ is unknown yet.
If there are no identical quarks in the final state, the remaining 
$\delta b_{ij}$'s vanish. Otherwise 
\begin{eqnarray}
\delta b_{33} \; = \; \delta b_{44} \; = \; \delta b_{34}, 
&& 
 \delta b_{55} \; = \;  \delta b_{66} \; \neq \;  0 
\qquad \quad
     \mbox{for  $b\rightarrow d \ov{d} d$\/ and $b\rightarrow s \ov{s} s$}
\end{eqnarray}
with $\delta b_{66}$ unknown.  The dependence on $\lambda_1$
parameterizing the effect of the b-quark's Fermi motion on the decay
rates cancels in the ratio in \eq{rqq}.

For the calculation of $\rnc$ in \eq{defrnc} we also need $r_{ue}$ and
$r_{sg}$. The corresponding expressions are \cite{hqe1,cfmrs}:
\begin{eqnarray}
r_{ue} &=& \lt| \frac{V_{ub}}{V_{cb}} \rt|^2 \frac{1}{f_1 (x_c^2) }
       \lt[ 1 + \frac{\as (\mu) }{2 \pi} 
                \lt[ h_{SL} (0) - h_{SL}(x_c) \rt]  + 
                6 \lt[ \frac{(1-x_c^2)^4}{f_1 (x_c^2) } -1 \rt] 
                 \frac{\lambda_2}{m_b^2} \rt] , \nn
r_{sg} &=& \lt| \frac{V_{tb}^* V_{ts}}{V_{cb}} \rt|^2 
           \frac{ 8 \, \as (\mu ) }{\pi \,f_1(x_c^2) } \lt[ C_8 (\mu) \rt]^2 
. \no
\end{eqnarray}

\section{Calculation}\label{sect:c}
This section is devoted to the calculation of the $g_{ij}$'s and
$b_{j8}$'s entering (\ref{peng},\ref{ga8},\ref{gaall}). All results 
correspond to the NDR scheme.

The first step is the same for all b-decays under consideration: 
The penguin diagram of \fig{fig:peng} must be calculated to obtain the
$r_{ij}$'s in \eq{r}:
\begin{eqnarray}
r_{24} \lt( p^2, m,\mu \rt) &=& 
     \frac{1}{3} \log \frac{m^2}{\mu^2} 
          - \frac{2}{9} - \frac{4 m^2}{3 p^2}   \nn 
&& -   \frac{1}{3} \lt( 1+ \frac{2 m^2}{p^2} \rt) 
            \sqrt{1 - \frac{4 m^2}{p^2} + i \delta } \,
            \log \frac{\sqrt{1 - \frac{4 m^2}{p^2} + i \delta} -1 }{
                       \sqrt{1 - \frac{4 m^2}{p^2} + i \delta} +1 } \;,   \nn
r_{24} \lt( p^2, 0 ,\mu \rt) &=& \frac{1}{3} \lt[ 
        \log \frac{p^2}{\mu^2} -i \pi \rt] - \frac{2}{9} \;,
\end{eqnarray}
\begin{eqnarray}
r_{26} &=& r_{24}, \qquad \qquad \qquad r_{23} \; = \; r_{25} \; = \; 
      - \frac{1}{3} r_{24} .
\end{eqnarray}
Here $p=p_b-p_d$ is the momentum flowing through the gluon
leg and $m$ is the internal quark mass. The infinitesimal ``$i
\delta$''-prescription yields the correct sign of the imaginary part
of the logarithm in the case $p^2 > 4 m^2$ and likewise regulates 
the square root for $p^2 < 4 m^2$. The $r_{1j}$'s and $r_{28}$ are zero.

Next we combine \eq{defgij} and \eq{r} to obtain the coefficients
$g_{ij}$ in \eq{peng} and \eq{gaall}:
\begin{eqnarray}
g_{ij} \lt( x_c, \frac{\mu}{m_b} \rt) 
   &=& \frac{16 \pi^3}{m_b^6} 
      \int \frac{d^3 \vec{p}_q d^3 \vec{p}_{\bar{q}} 
        d^3 \vec{p}_d}{(2\pi)^5 8 |E_q E_{\bar{q}} E_d| }  
   \delta^{(4)} \left( p_b + p_{\bar{q}} - p_q -p_d \right) \cdot \nn 
&& \hspace{17ex}
   \sum_{k=1}^6 r^c_{ik} \lt( (p_b-p_d)^2, x_c m_b, \mu \rt)
   \ov{ \langle Q_{k} \rangle^{(0)} \langle Q_{j} \rangle^{(0)\,*} } .
\label{gijr}
\end{eqnarray}
The corresponding expression for an internal $u$-quark is obtained by
substituting $c$ with $u$ in \eq{gijr}.  For the decays $b\rightarrow
u \ov{u} d$, $b\rightarrow u \ov{u} s$, $b\rightarrow s \ov{s} d$ and
$b\rightarrow d \ov{d} s$ one finds:
\begin{eqnarray}
g_{22} \lt( x_c, \frac{\mu}{m_b} \rt) & = & g_{24} \lt( x_c,
 \frac{\mu}{m_b}  \rt)  \; = \; 
g_{26} \lt( x_c, \frac{\mu}{m_b}  \rt) \; =  \nn 
&& \hspace{-45pt}
\frac{16}{27} \ln \frac{x_c m_b}{\mu } - \frac{16}{27} 
 \lt( 1- 10 x_c^2 + 18 x_c^4 -36 x_c^6 \rt) \sqrt{1- 4 x_c^2} \,
\ln \frac{1-\sqrt{1- 4 x_c^2} }{2 x_c} \no \\[0.3ex] 
&& \hspace{-45pt}
+ \frac{4}{9} \lt[ - \frac{3}{2} + 16 x_c^2 - 14 x_c^4 + 24 x_c^6  + 
    32 x_c^6 \lt( 2- 3 x_c^2 \rt) 
       \lt( \ln^2  \frac{1-\sqrt{1- 4 x_c^2} }{2 x_c} -
             \frac{\pi^2}{4} \rt)  \rt] \no \\[0.3ex]
&& \hspace{-45pt}
- i \, \pi  \frac{4}{9} \lt[\frac{2}{3}  \sqrt{1- 4 x_c^2} 
     \lt( 1- 10 x_c^2 + 18 x_c^4 -36 x_c^6  \rt)  - 32 x_c^6 \lt (2- 3 x_c^2
      \rt) \ln  \frac{1-\sqrt{1- 4 x_c^2} }{2 x_c} \rt] \no \\[1ex]
g_{22} \lt( 0, \frac{\mu}{m_b} \rt) & = & g_{24} \lt( 0 ,
 \frac{\mu}{m_b}  \rt)  \; = \; 
  g_{26} \lt( 0, \frac{\mu}{m_b}  \rt) \; = \;
 \frac{4}{9} \lt[ - \frac{3}{2} + \frac{4}{3} \ln \frac{m_b}{\mu} 
     - \frac{2}{3} i \pi \rt] \no \\[1ex] 
g_{21} \lt( x, \frac{\mu}{m_b} \rt)  & = & 
g_{23} \lt( x, \frac{\mu}{m_b} \rt) \; = \; 
g_{25} \lt( x, \frac{\mu}{m_b} \rt) \; = \;  
g_{1j} \lt( x, \frac{\mu}{m_b} \rt) \; = \; 0,  
\qquad j=1,\ldots 6, \label{g22}
\end{eqnarray} 
where $x_c=m_c/m_b$. Numerically one finds for actual quark masses:
\begin{eqnarray}
g_{22} \lt( 0, 1 \rt) &=& -0.67 - 0.93 \, i \, ,  \qquad \qquad 
g_{22} \lt( 0.3, 1 \rt) \; = \; -0.69 - 0.23 \, i \, . \label{numex}
\end{eqnarray}
The near equality of the real parts in \eq{numex} is a numerical
accident. 

In the case of two identical particles in the final state both
diagrams of \fig{fig:pinc} contribute.  Then $g_{23}$ is no more zero,
but
\begin{eqnarray}
g_{23} \lt( x, \frac{\mu}{m_b} \rt) &=& g_{22} \lt( x, \frac{\mu}{m_b}
\rt) \qquad\qquad
    \mbox{for  $b\rightarrow d \ov{d} d$\/ and $b\rightarrow s \ov{s} s$.}
\label{g23}
\end{eqnarray} 
The remaining $g_{ij}$'s are as in \eq{g22}. As an analytical check we
have confirmed that the $\mu$-depen\-dence in \eq{g22} and \eq{g23}
cancels with the $\mu$-dependence in the Wilson coefficients in
\eq{gtree} to order $\as$.

We now turn to the calculation of $\Delta \Gamma_8$ in \eq{ga8}.
Performing the phase space integration in \eq{defbj8} yields 
\begin{eqnarray}
b_{28} &=&  b_{48} \; = \; b_{68} \; = \; - \frac{16}{9} \; , \nn
b_{18} &=&  b_{38} \; = \; b_{58}  \; = 0 \label{res8}
\end{eqnarray}
for $b\rightarrow u \ov{u} d$, $b\rightarrow u \ov{u} s$,
$b\rightarrow s \ov{s} d$ and $b\rightarrow d \ov{d} s$.
If identical particles are present, $b_{38}$ is no more zero,
but instead reads
 \begin{eqnarray}
b_{38} &=& - \frac{16}{9}  \qquad \qquad
    \mbox{for  $b\rightarrow d \ov{d} d$\/ and $b\rightarrow s \ov{s} s$.}
\end{eqnarray}
The remaining $b_{j8}$'s are as in \eq{res8}.

\section{Charmless decay rates}\label{sect:num}
\subsection{Standard Model}
In this section we discuss our numerical results for the various decay
rates. We use the following set of input parameters: 
\begin{eqnarray}
&& \hspace{-1ex}
\begin{array}[b]{llll}
\displaystyle
\lt| \frac{V_{ub}}{V_{cb}} \rt| = 0.08\pm 0.02 , \qquad  
& \displaystyle
\delta = 90^\circ \pm 30^\circ ,
& \displaystyle
x_c = 0.29 \pm 0.04 ,
& \displaystyle 
\mu = m_b = 4.8~\mbox{GeV} , \\[4mm]
\displaystyle 
\as \lt( M_Z \rt) = 0.118 ,
& \displaystyle 
\lt| V_{cb} \rt| = 0.038 , 
& \displaystyle 
m_t \lt( m_t \rt) = 168~\mbox{GeV} . 
& 
\end{array}
\label{inp}
\end{eqnarray}
The $\rqq{q}{q}{q^\prime}$'s sizeably depend on $|V_{ub}/V_{cb}|$,
$\delta$, $x_c$ and especially on the renormalization scale, which
will be varied in the range $m_b/2 \leq \mu \leq 2 m_b $. The quark
masses in \eq{inp} are taken from \cite{bsu}, the values for
$|V_{ub}/V_{cb}|$ and $|V_{cb}|$ have been presented in \cite{a}.
The range for the CKM phase $\delta$ has been obtained from the NLO  
analysis of $\epsilon_K$ and $\Delta m_{B}$ in \cite{hn3}. The 
dependence on $\as \lt( M_Z \rt)$ in the range 
$0.112 \leq \as \lt( M_Z \rt) \leq 0.124$ \cite{beth} is weaker than
the $\mu$-dependence. Our results are listed in \tab{tab:res}.

\begin{ntable}{tb}
\begin{displaymath}
\begin{array}{r||c|l|l|l|l|l|l|l}
 & & \multicolumn{7}{c}{\rule[-3mm]{0pt}{6mm} \mbox{final state}} 
                     \\ \rule[-3mm]{0pt}{8mm}   
 & \mbox{input}    & u \ov{u} d & u \ov{u} s & d \ov{d} s & 
                     s \ov{s} s & s \ov{s} d & d \ov{d} d & 
                     \mbox{no charm}
                     \\\hline\hline \rule[-3mm]{0pt}{8mm}
\displaystyle 
\rqq{q}{q}{q^\prime} & \mbox{as in  \eq{inp}}  
                     & 0.040 & 0.021 & 0.018 & 0.015 & 8.9 \cdot
                     10^{-4} & 7.2 \cdot 10^{-4}  
                     & 0.14
                  \\ \cline{2-9} \rule[-3mm]{0pt}{8mm} 
& \mu=m_b/2          & 0.044 & 0.033 & 0.029 & 0.024 & 14.0 \cdot
                     10^{-4} & 11.4 \cdot 10^{-4} 
                     & 0.19 
                     \\ \cline{2-9} \rule[-3mm]{0pt}{8mm} 
& \mu = 2 \, m_b        & 0.037 & 0.014 & 0.011 & 0.009 & 5.5 \cdot
                     10^{-4} & 4.6 \cdot 10^{-4} 
                     & 0.11 
                     \\ \cline{2-9} \rule[-3mm]{0pt}{8mm} 
& \lt| \frac{V_{ub}}{V_{cb}} \rt| = 0.06   
                     & 0.023 & 0.020 & 0.018 & 0.015 & 8.7 \cdot
                     10^{-4} & 7.1 \cdot 10^{-4} 
                     & 0.11
                     \\ \cline{2-9} \rule[-3mm]{0pt}{8mm} 
& \lt| \frac{V_{ub}}{V_{cb}} \rt| = 0.10   
                     & 0.062 & 0.023 & 0.018 & 0.015 & 9.1 \cdot
                     10^{-4} & 7.5 \cdot 10^{-4} 
                     & 0.18 
                     \\ \cline{2-9} \rule[-3mm]{0pt}{8mm}                      
& \delta  = 60^\circ   
                     & 0.044 & 0.017 & 0.018 & 0.015 & 5.9 \cdot
                     10^{-4} & 4.8 \cdot 10^{-4} 
                     & 0.14
                     \\ \cline{2-9} \rule[-3mm]{0pt}{8mm}                      
& \delta  = 120^\circ  
                     & 0.036 & 0.025 & 0.018 & 0.014 & 12.2 \cdot
                     10^{-4} & 10.0 \cdot 10^{-4} 
                     & 0.14 
                     \\ \cline{2-9} \rule[-3mm]{0pt}{8mm}                      
& x_c = 0.25 
                     & 0.034 & 0.019 & 0.016 & 0.013 & 8.1 \cdot
                     10^{-4} & 6.7 \cdot 10^{-4} 
                     &0.12 
                     \\ \cline{2-9} \rule[-3mm]{0pt}{8mm}                      
& x_c = 0.33  
                     & 0.048 & 0.023 & 0.020 & 0.017 & 9.7 \cdot
                     10^{-4} & 7.9 \cdot 10^{-4} 
                     & 0.16
                     \\ \hline\hline \rule[-3mm]{0pt}{10mm}     
\displaystyle 
Br                   & \mbox{as in  \eq{inp}}  
                     & 0.41\% & 0.22\% & 0.18\% & 0.15\% 
                     & 9.1 \cdot 10^{-3}\% & 7.4 \cdot 10^{-3}\% 
                     & 1.4 \%
\end{array}
\end{displaymath}
\caption{ The values of $\rqq{q}{q}{q^\prime}$ for the various 
  final states as defined in \eq{defr}. The input parameters are
  chosen as in \eq{inp} except for the quantity listed in the second
  column.  The last column lists $\rnc$ defined in \eq{defrnc}. $Br$
  in the last row is the branching ratio for $B \rightarrow 
  X_{q\ov{q}q^\prime}$ obtained by multiplying $ \rqq{q}{q}{q^\prime}$
  with $B_{SL} = 10.23\% $.
}\label{tab:res} 
\end{ntable} 

Keeping the physical input parameters as in \eq{inp} and varying the 
scale in the range $m_b/2 \leq \mu \leq 2 m_b $ the charmless decay
modes sum to 
\begin{eqnarray}
\rnc &=& 0.15 \pm 0.04 .\no 
\end{eqnarray}
The values for $r_{sg}$ and $r_{ue}$ entering this result are 
\begin{eqnarray}
r_{sg} \; = \; 0.02 \pm 0.01, \qquad && \qquad  
r_{ue} \; = \; 0.01 \pm 0.00 . \no
\end{eqnarray}
Incorporating also the uncertainties in $|V_{ub}/V_{cb}|$ and $x_c$ 
one finds
\begin{eqnarray}
\rnc &=& 0.15 \pm 0.08 . \label{rcres} 
\end{eqnarray}

To discuss the results for the individual charmless decay modes it 
is instructive to look at the separate scheme-independent 
contributions from $\Gamma^{(0)}$, $\Delta \ov{\Gamma}_\mathit{cc}$, 
$\Delta \ov{\Gamma}_\mathit{peng}$, $\Delta \ov{\Gamma}_8$ and 
$\Delta \ov{\Gamma}_\mathit{W}$ (cf.\ the appendix) to the decay rate.
These contributions are listed in \tab{tab:sep}, in which also 
the contributions from penguin operators to $\Gamma^{(0)}$ are shown.

\begin{ntable}{tb}
\begin{displaymath}
\begin{array}{r||r|r|r|r|r|r}
\rule[-3mm]{0pt}{8mm}     
\mbox{final state} & \propto C_{1,2}^{(0)} \cdot C_{1,2}^{(0)} &
                     \propto C_{3-6}^{(0)} \cdot C_{1-6}^{(0)} &
                     \propto \Delta \ov{\Gamma}_\mathit{W} &
                     \propto \Delta \ov{\Gamma}_\mathit{cc} & 
                     \propto \Delta \ov{\Gamma}_\mathit{peng} &
                     \propto \Delta \ov{\Gamma}_8 
                     \\ \hline\hline \rule[-3mm]{0pt}{8mm}     
 u \ov{u} d & 107 & -4 & -5 &  6 & - 5   & 1 
                     \\ \hline\hline \rule[-3mm]{0pt}{8mm}     
 u \ov{u} s & 10 & 39 & -7 & 1 & 72 & -14 
                     \\ \hline\hline \rule[-3mm]{0pt}{8mm}     
 d \ov{d} s & 0 & 46 & -8 & 0 & 78 & -16  
                     \\ \hline\hline \rule[-3mm]{0pt}{8mm}     
 s \ov{s} s & 0 & 44 &  -20 & 0 &  92 &  -16
                     \\ \hline\hline \rule[-3mm]{0pt}{8mm}     
 s \ov{s} d & 0 & 54 & -9 & 0 & 74  & -19
                     \\ \hline\hline \rule[-3mm]{0pt}{8mm}     
 d \ov{d} d & 0 & 52 & -24& 0 & 91 & -19 
                     \\ \hline\hline \rule[-3mm]{0pt}{8mm}     
 s g        & 0 &  0 &   0& 0 &  0 & 100  
                     \\ \hline\hline \rule[-3mm]{0pt}{8mm}     
 \mbox{no charm}  & 46 &  17 &  -6 &  2 &  31 &  11
\end{array}
\end{displaymath}
\caption{Separate contributions to 
  $\Gamma (b\rightarrow q \ov{q} q^\prime)$ in percent of the total
  rate for the input parameters of \eq{inp}.  In the third column the
  contribution from the LO matrix elements of penguin operators and
  their interference with matrix elements of $Q_{1,2}$ is shown. In
  the last row $\Gamma (b\rightarrow u e \ov{\nu}_e) $ entering
  $\rnc$ has been assigned to the second column listing 
  the current-current part.  }\label{tab:sep}
\end{ntable}
All decays except for $B \rightarrow X_{u\ov{u}d}$ are dominated by
penguin effects. The sizeable $\mu$-dependence in these decays can be 
reduced by calculating the current-current type radiative corrections to
penguin operators. Note that all these decay rates are even dominated 
by the \textit{penguin diagrams}\ calculated in this work. On the 
other hand the terms stemming from $Q_8$ lower the penguin induced rates.
In $B \rightarrow \textit{no charm}$, however, the net effect 
of $Q_8$ is positive because of the two-body decay  $b \rightarrow
s g$ proportional to $C_8^2$. 

The calculation of $\rnc$ in \cite{ap} has included $\Gamma^{(0)}$,
$\Delta \ov{\Gamma}_\mathit{W}$, and $\Delta \ov{\Gamma}_\mathit{cc}$.
Taking into account that in \cite{ap} a high (theoretical) value for
$B_{SL}$ has been used, the result of \cite{ap} translates into 
\begin{eqnarray}
\rnc &=& 0.14 \qquad \mbox{for } \lt| \frac{V_{ub}}{V_{cb}} \rt| =
0.10 .\no
\end{eqnarray}
The corresponding value in \tab{tab:res} is $\rnc=0.18$ showing the
increase due to $\Delta \ov{\Gamma}_\mathit{peng} + \Delta
\ov{\Gamma}_8$.

Despite of the 36 \% increase in $\rnc$ in \eq{rcres} the value is
still much below $\rnc = 1.0\pm 0.4$ needed to solve the missing charm
puzzle. The new theoretical prediction for $n_c$ is 
\begin{eqnarray}
n_c &=& 1.33 \mp 0.06, \no
\end{eqnarray}
which is only marginally lower than the old result in \eq{puzz}.
Still an enhancement of $\rnc$ by a factor between 2.6 and 20 is
required.  Yet with the result in \eq{g22} for the penguin diagram of
\fig{fig:pinc} at hand we can estimate the non-perturbative
enhancement of the $(c,\ov{c})$ intermediate state necessary  to
reproduce the experimental result. The mechanism proposed in 
\cite{disy,ps} corresponds to a violation of quark-hadron duality,
which we may parametrize by multiplying the $(c,\ov{c})$-penguin  
function $g_{ij} \lt( x_c, \mu/m_b \rt)$ in \eq{gaall} 
by an arbitrary factor $d$. We find that $d$ must be chosen as large
as 20 in order to reach the lowest desired value $\rnc=0.6$ for the 
central set of the input parameters in \eq{inp}. Yet in this case one 
must also include the double-penguin diagram obtained by squaring 
the result of \fig{fig:peng}. Taking this into account 
non-perturbative effects must still increase the $(c,\ov{c})$-penguin
diagram by roughly a factor of 9 over its short distance 
result in the NDR scheme. Having in mind that the phase space
integration contained in  $g_{ij} \lt( x_c, \mu/m_b \rt)$ 
implies a smearing of the invariant hadronic mass of the 
$(c,\ov{c})$-pair such a large deviation of quark-hadron duality 
seems unlikely. 

Our results for the branching fractions of $b\rightarrow d\ov{d} s$
and $b\rightarrow s\ov{s} s$ differ by roughly a factor of 1/2 from
the results in \cite{f}. The main source of the discrepancy is the
different calculation of the total rate $\Gamma_{tot}$ entering the
theoretical definition of the branching fraction. Our predictions for
the branching ratios in \tab{tab:res} contain the total rate via
$\Gamma_{tot}=\Gamma_{SL}/B^\mathit{exp}_{SL}$, while in \cite{f}
$\Gamma_{tot}=G_F^2 m_b^5 |V_{cb}|^2 /(64 \pi^3) $ has been used. This
approximation neglecting the RG effects appears to be too crude and is
responsible for 50 \% of the discrepancy. The remaining difference is
due to the effect of $\Delta \Gamma_8$ not considered in \cite{f} and
the use of different values of the Wilson coefficients. Our
predictions for $Br (b\rightarrow s\ov{s} d)$ and $Br (b\rightarrow
d\ov{d} d)$ are even smaller by a factor of 5 than the results of
\cite{f}. This is due to the fact that in addition the value of
$|V_{td}/V_{cb}|^2$ used in \cite{f} is more than twice as large as
the present day value used in our analysis. We remark that with our
new results the number of B-mesons to be produced in order to detect
the inclusive CP asymmetries corresponding to these decays is
substantially larger than estimated in \cite{f}.

\subsection{New physics} 
In the Standard Model the initial conditions for $C_{3-6}$ and $C_{8}$
are generated at a scale $\mu=O(M_W)$ by the one-loop $bsg$-vertex
function with a $W$-boson and a top quark as internal particles.  Due
to the helicity structure of the couplings positive powers of $m_t$
are absent in the $bsg$-vertex. In many extensions of the Standard
model such a helicity suppression does not occur \cite{k}.  Further in
supersymmetric extensions flavour changing transitions can be mediated
by gluinos, whose coupling to (s)quarks involves the strong rather the
weak coupling constant. Recently a possible enhancement of $C_8$
affecting $r_{sg}$ and thereby $\rnc$ has attracted much attention
\cite{k}. In the following we will perform a model independent
analysis of this hypothesis.

New physics  affects the initial condition of the Wilson coefficients
calculated at the scale of the masses mediating the flavour-changing
transition of interest. We will take $\mu=M_W$ as the initial scale 
for both the standard and non-standard contributions to the Wilson
coefficients. The renormalization group evolution down to $\mu=m_b$
mixes the various initial values and 
can damp or enhance the new physics effects in $C_i (M_W)$. For
example the Standard Model value of $C_8(4.8 \,\textrm{GeV})$ in
\tab{tab:wc} is mainly a linear combination of $C_2(M_W)$ and
$C_8(M_W)$:
\begin{eqnarray}
C_8 \lt( m_b \rt) &=& -0.08\, C_2 \lt( M_W \rt) \, +\, 0.7 \,C_8 \lt(M_W\rt)
\no
\end{eqnarray}
with $C_8 (M_W) =-0.1$. 
\begin{nfigure}{tb}
\vspace{-10truemm}
\centerline{\epsfxsize=0.8\textwidth \rotate[r]{\epsffile{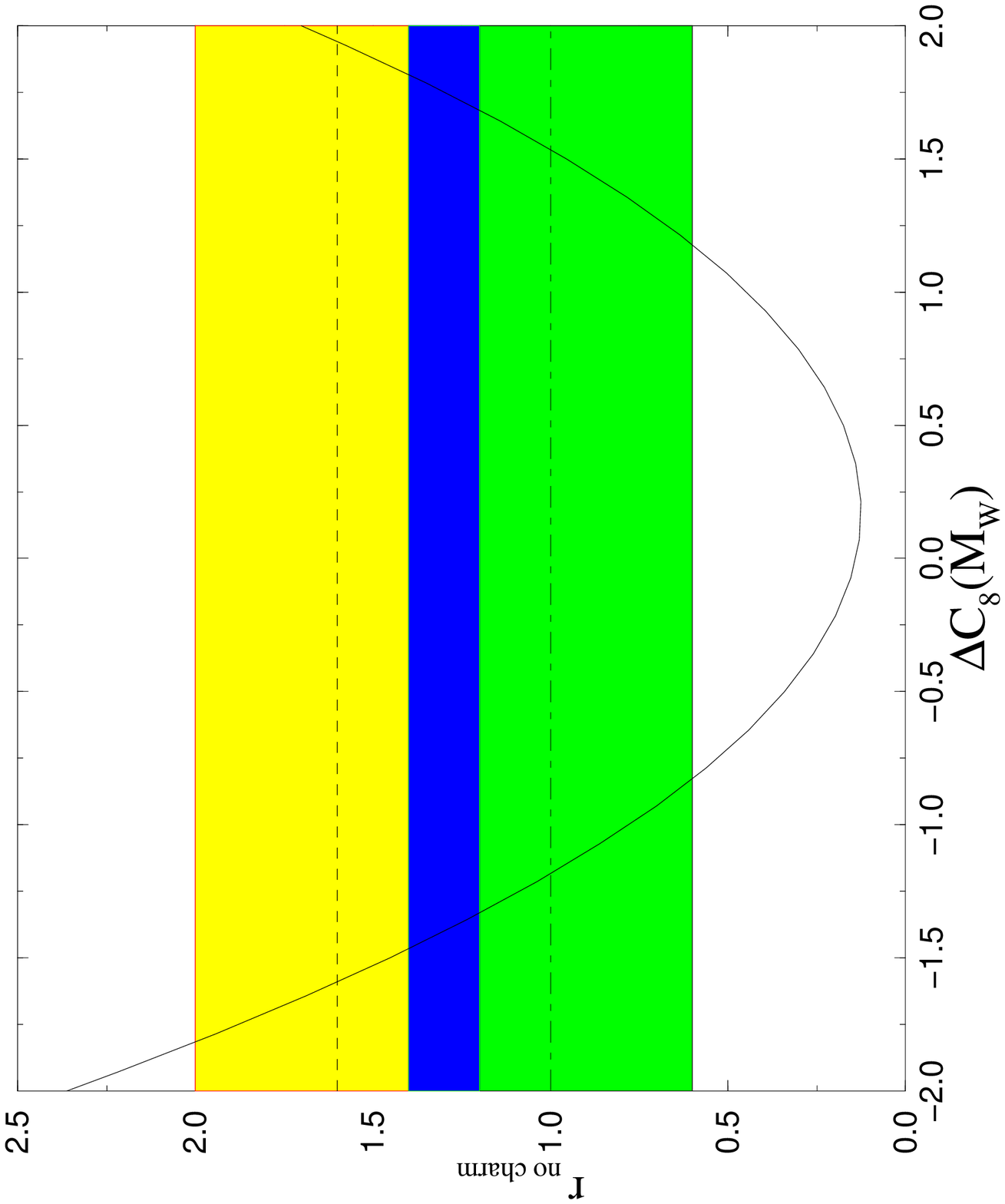}}}
\caption{$\rnc$ vs.\ $\Delta C_8 (M_W)$ parameterizing new physics 
contributions to $C_8 (M_W)$. The dark shading  marks the region 
$\rnc =1.0 \pm 0.4$ needed to reproduce the experimental result for 
$n_c$ in \eq{nc}. The lightly shaded area corresponds to  
$\rnc =1.6 \pm 0.4$ obtained from the analysis in \cite{disy}.
}
\label{fig:new}
\end{nfigure}
If one enlarges $C_8 (M_W)$ by a factor of 10 while keeping
$C_{1-6}(M_W)$ fixed, $\rnc$ grows by a factor of 5.5. The sensitivity
of $\rnc$ to $C_{3-6}(M_W)$ is much smaller, increasing the latter by
a factor of 10 enhances $\rnc$ only by a factor of 1.6. Hence in the
following we will only focus on $C_8 \lt(M_W \rt)=-0.1+\Delta C_8
(M_W)$, where $\Delta C_8 (M_W)$ is the new physics contribution. For
simplicity we will further assume that the CKM structure of the new
contributions is the same as in the Standard model and neglect the
possibility of new CP-violating phases by assuming $\Delta C_8 (M_W)$
to be real.

In \fig{fig:new} $\rnc$ is plotted versus $\Delta C_8 (M_W)$. Solving
for $\rnc = 1.0\pm 0.4$ yields two solutions:
\begin{eqnarray}
\Delta C_8 (M_W) \; = \; -1.2 
\begin{array}[c]{l} \scriptstyle  -0.3 \\[-2mm] \scriptstyle +0.4 \end{array}
 \, ,  \qquad &&
\qquad
\Delta C_8 (M_W) \; = \; 1.5 \pm 0.3 \, . \label{npsol}
\end{eqnarray}
The central values correspond to an enhancement of the Standard Model
value for $C_8(M_W)$ by factors of 13 and $(-14)$. We hope to resolve 
the twofold ambiguity after calculating the contribution of $Q_8$ to
$\Gamma (\ov{B} \rightarrow X_{ c \ov{c} s }) $.

The LEP data in \eq{lepz} correspond to $\rnc = 0.4 \pm 0.5$ and 
\begin{eqnarray}
\Delta C_8 (M_W) \; = \; -0.5  
\begin{array}[c]{l} \scriptstyle  -0.6 \\[-2mm] \scriptstyle +0.5 \end{array}
 \, ,  \qquad &&
\qquad
\Delta C_8 (M_W) \; = \; 0.9 
\begin{array}[c]{l} \scriptstyle  +0.6 \\[-2mm] \scriptstyle -0.9 \end{array}
\no
\end{eqnarray}
showing the consistency of \eq{lepz} with the Standard Model.  

\section{Conclusions}
We have calculated two new contributions to the inclusive decay rates
of B-mesons into various charmless final states: First we have
obtained the results of penguin diagrams involving the operator $Q_2$
and a $c$- or $u$-quark in the loop putting special care on the
renormalization scheme independence of our results. Second we have
calculated the influence of the chromomagnetic dipole operator $Q_8$
on these decays.  The former contributions have been found to dominate
the branching fractions for $\ov{B}\rightarrow X_{u \ov{u} s}$,
$\ov{B}\rightarrow X_{d \ov{d} s}$, $\ov{B}\rightarrow X_{s \ov{s}
  s}$, $\ov{B}\rightarrow X_{s \ov{s} d}$ and $\ov{B}\rightarrow X_{d
  \ov{d} d}$.  The effect of $Q_8$ on these decay modes is also
sizeable and decreases the rates. On the other hand the decay rate for
$\ov{B}\rightarrow X_{u \ov{u} d}$ is only affected by a few percent.
Our results increase the theoretical prediction for $ Br ( B
\rightarrow \textit{no charm} )$ by 36 \%, which is not sufficient to
explain the charm deficit observed in B-decays by ARGUS and CLEO. If a
breakdown of quark-hadron duality due to intermediate $(c,\ov{c})$
resonances is to explain the ``missing charm puzzle'', the phase space
integrated penguin diagram with an internal $c$-quark must be larger
than the perturbative result in the NDR scheme by roughly a factor of
9.

We have then analyzed the hypothesis that new physics effects enhance
the coefficient $C_8$ of $Q_8$ and have performed a model independent
fit of $C_8$ to the experimental data on $n_c$. The renormalization
group evolution from $\mu=M_W$ to $\mu=m_b$ has been properly taken
into account. One finds two solutions for $C_8$: For the central
values of the theoretical input and the data of the $\Upsilon (4 S)$
experiments $C_8(M_W)$ must be larger by a factor of 13 or $(-14)$
than in the Standard Model, if the new physics contributions have the
same CKM structure as the Standard Model penguin diagram.

\section*{Acknowledgements} 
We are grateful to Andrzej Buras for many stimulating discussions.  
We thank him and Christoph Greub for proofreading the manuscript. 
U.N.\ acknowledges interesting discussions with Stefano Bertolini, 
Antonio Masiero and Yong-Yeon Keum.

\appendix 
\section{Scheme independence}
Now we discuss the cancellation of scheme dependent terms between the
NLO Wilson coefficients and the loop diagrams contained in $\Delta
\Gamma_{cc}$ and $\Delta \Gamma_{peng}$. We define scheme independent
combinations $\Delta \ov{\Gamma}_{cc}$, $\Delta \ov{\Gamma}_{peng}$
and $\Delta \ov{\Gamma}_W$, which allow a meaningful discussion of the
numerical sizes of these separate contributions to $\Gamma$ as
performed in sect.~\ref{sect:num}. Finally we comment on the scheme
independence of $\Delta \Gamma_8$.

The NLO correction to the Wilson coefficients in \eq{wc} can be split
into two parts:
\begin{eqnarray}
\Delta C_j (\mu) &=&  
       \sum_{k=1}^6 J_{jk} C_k^{(0)} (\mu) + \Delta \ov{C}_j (\mu) 
, \qquad\qquad j=1,\ldots 6. \label{wca}
\end{eqnarray} 
$\Delta \ov{C}_j (\mu)$ contains the contributions stemming from the
weak scale. It is independent of the renormalization scheme and
proportional to $\as (M_W)/\as(\mu)$.  Yet the $J_{jk}$'s in \eq{wca}
are scheme dependent. The precise definitions of the terms in \eq{wca}
can be found in \cite{bjlw,hn}.  We now absorb the terms involving
$J_{jk}$ into $\Delta \ov{\Gamma}_\mathit{cc}$ and $\Delta
\ov{\Gamma}_\mathit{peng}$, so that the latter become scheme
independent.

The identification of scheme independent combinations of one-loop
matrix elements and $J_{jk}$'s is most easily done, if one expresses
the loop diagrams in terms of the tree-level matrix elements.  The
combination
\begin{eqnarray}
  r^{q^\prime}_{jk} \lt( p^2, m_{q^\prime},\mu \rt) 
     &+& J_{kj},  \qquad \qquad \mbox{$j\leq 2$ and $k\geq 3$,} \label{rpJ}
\end{eqnarray}
of the coefficients in \eq{r} and the $J_{kj}$'s is scheme independent 
\cite{bjlw}. Substituting $r^{q^\prime}_{jk} $ with \eq{rpJ} in 
\eq{gijr} one finds the scheme independent quantity:
\begin{eqnarray} 
\Delta \ov{\Gamma}_\mathit{peng} &=&   \Delta \Gamma_\mathit{peng} +
\frac{G_F^2 m_b^5}{64 \pi^3} 
   \, 2 \,  \real    \lt[ - \, t \,   \xi_u \xi_t^*   
    \sum_{ \scriptstyle i,j=1,2 \atop \scriptstyle 
            k=3,\dots 6} C_i^{(0)} C_j^{(0)} \,
     J_{ki} b_{jk}   \rt. \nn
&& \phantom{   \Delta \Gamma_\mathit{peng} + \frac{G_F^2 m_b^5}{64 \pi^3} 
                \, 2 \,  \real   \xi_t^*  [ }  \;
\lt. +    \lt| \xi_t \rt|^2 \! 
  \sum_{ \scriptstyle i = 1,2 \atop \scriptstyle 
         j,k  = 3, \ldots 6} \! C_i^{(0)}  C_j^{(0)} 
    J_{ki} b_{jk}   \rt] . \label{penga}
\end{eqnarray} 
Here  $\xi_t^* = - \xi_u^* -\xi_c^* $ has been used.
In the NDR scheme the $J_{ki}$'s in \eq{penga} evaluate to
\cite{bjlw}
\begin{eqnarray}
  J_{31} 
=-0.877, && \quad 
J_{32}^\mathit{NDR} = 
-0.532, \no\\[0.3ex] 
J_{41} =0.324, && \quad 
J_{42}^\mathit{NDR} = -0.202,
\no\\[0.3ex] 
J_{51} =0.557,&& \quad  
J_{52}^\mathit{NDR} = 
0.511,\no\\[0.3ex] 
J_{61} =0.146, && \quad 
J_{62}^\mathit{NDR} = -0.677 \, . \label{j}
\end{eqnarray}
The $J_{k1}$'s in the first row of \eq{j} do not depend on the
renormalization scheme, because $r^{q^\prime}_{1j}=0$ in all schemes
due to a vanishing colour factor.

In the same way one finds 
\begin{eqnarray}
\Delta \ov{\Gamma}_{cc} &=& t \,  \frac{G_F^2 m_b^5}{64 \pi^3} 
   \, 2 \, \left| \xi_u \right|^2 \sum_{i,j=1}^2 
   C_i^{(0)} C_j^{(0)} \lt[ h_{ij} + \sum_{k=1}^2 J_{ki} b_{kj} \rt].
\label{cca}
\end{eqnarray}
Here the scheme dependence of the $h_{ij}$'s in \eq{defh} cancels with
the one of the $J_{ki}$'s \cite{bjlw}:
\begin{eqnarray}
J_{11}^\mathit{NDR} = J_{22}^\mathit{NDR} = 
\frac{631}{6348} = 0.099 ,&& \qquad 
J_{12}^\mathit{NDR} = J_{21}^\mathit{NDR} = \frac{3233}{2116} =  1.528 .
\label{jcc}
\end{eqnarray}
If one inserts \eq{wca} into $\Delta \Gamma_W$, one finds the
$J_{jk}$'s with $k\leq 2$ to appear exactly in the combinations entering
\eq{penga} and \eq{cca}. The remaining $J_{jk}$'s with $k \geq 3$
describing penguin-penguin mixing would cancel the scheme dependence of 
the loop diagrams of \fig{fig:cc} and \fig{fig:peng} with insertions
of penguin operators $Q_j$, $j=3,\ldots 6$. Since the latter are
omitted in our calculation, we must also leave out the $J_{jk}$'s with
$k \geq 3$ in \eq{wca}. This has been done in \tab{tab:wc}. 
$\Delta \ov{C}_j $ has been tabulated in the last line of \tab{tab:wc}
for illustration.  It can be obtained from the other entries of the
table with the help of \eq{wca}, \eq{jcc} and \eq{j}.

Finally $\Delta \ov{\Gamma}_{W}$ is simply obtained from $\Delta
\Gamma_W$ in \eq{gw} by replacing $\Delta C_j$ with $\Delta \ov{C}_j$.

Unlike $C_j^{(0)}$, $j \leq 6 $, $C_8^{(0)}$ is a two-loop quantity
and therefore a priori scheme dependent.  We understand $H$ in \eq{hd}
to be renormalized such that the matrix elements $\bra{d g}Q_j\ket{b}$
vanish at the one-loop level for $j=1,\ldots 6$. This ensures that
$\Delta \Gamma_8$ as defined in \eq{ga8} is scheme independent
\cite{cfmrs}. The thereby renormalized LO coefficient $C_8^{(0)}$ ist
usually called $\widetilde{C}_8$ or $C_8^{(0),eff}$. In the NDR scheme
this finite renormalization simply amounts to $C_8^{(0)} =
C_8^{(0),NDR}+C_5^{(0)}$. Apart from $\Delta \Gamma_8$ this only affects
the penguin diagram of $Q_5$ (cf.~\fig{fig:peng}), which is a part of
the neglected radiative corrections to penguin operators.


\begin{thebibliography}{99}
\bibitem{br} T.E.~Browder, K.~Honscheid and D.~Pedrini,
  hep-ph/9606354, to appear in 
  \textit{Annual Review of Nuclear and Particle Science}. \\ 
  T.E.~Browder, hep-ph/9611373, 
  talk at the \textit{ICHEP conference}\ 1996, Warsaw, 
   to appear in the proceedings. \\
  J.D.~Richman, hep-ex/9701014, talk at the \textit{ICHEP conference},
  Warsaw, 1996. 
\bibitem{ex} 
  B.~Barish \textit{et al.}\/ (CLEO), \prl 76 (1996) 1570.\\
  H.~Albrecht \textit{et al.}\/ (ARGUS), \plb 318 (1993) 397.
\bibitem{hqe1}
I.I.~Bigi, N. Uraltsev, and A. Vainshtein, Phys.\ Lett.\ B 293,
430 (1992); Erratum \textit{ibid.} 297, 477 (1993). \\
B.~Blok and M.~Shifman, \npb 399 (1993) 441; \textit{ibid.} 399
  (1993) 459. 
\bibitem{hqe2}
A. Manohar and M. Wise, Phys.\ Rev.\ D 49, (1994) 1310.\\
B. Blok, L.~Koyrakh, M. Shifman and A.I. Vainshtein, 
\prd 49 (1994), 3356; Erratum \textit{ibid.} D50 (1994) 3572. \\ 
T. Mannel, \npb 413 (1994) 396.\\
I.I.~Bigi, M.A. Shifman, N.G.~Uraltsev, A.I.~Vainshtein,
Int.~J.\ Mod.\ Phys.\ A9 (1994) 2467.
\bibitem{spec} I.I.~Bigi, B.~Blok, M.~Shifman, N.~Uraltsev and 
               A.~Vainshtein, in \textit{B decays}, ed. S.~Stone, 
               2nd edition, \textit{World Scientific}, Singapore,
               1994, 132.\\
               I.I.~Bigi, hep-ph/9508408.
\bibitem{ns}   M.~Neubert and C.~Sachrajda, hep-ph/9603202. 
\bibitem{kroll} I.J.~Kroll, hep-ex/9602005, 
 proceedings of the  
 \textit{17th Int.\ Symp.\  on Lepton Photon Interactions}, 
  Beijing, P.R.~China, 1995, 204.
\bibitem{bbbg1} E.~Bagan, P.~Ball, V.M.~Braun and P.~Gosdzinsky,
         \npb 432 (1994) 3.
\bibitem{bbfg} E.~Bagan, P.~Ball, B.~Fiol and P.~Gosdzinsky,
\plb 351 (1995) 546. 
\bibitem{bbbg2} E.~Bagan, P.~Ball, V.M.~Braun and P.~Gosdzinsky,
\plb 342 (1995) 362; Erratum \textit{ibid} B374 (1996) 363.
\bibitem{ap} G.~Altarelli and S.~Petrarca, \plb 261 (1991) 303. 
\bibitem{n} M.~Neubert, hep-ph/9605256, to appear in the proceedings
  of the \textit{10th Les Rencontres de Physique de la Vallee d'Aoste},
  La Thuile, 1996.
\bibitem{bdy} 
  G.~Buchalla, I.~Dunietz and H.~Yamamoto,
  \plb 364 (1995) 188.
\bibitem{disy} 
  I.~Dunietz, J.~Incandela, F.D.~Snider and H.~Yamamoto, hep-ph/9612421.
\bibitem{lep} ALEPH coll., EPS-404, contribution to the \textit{Int.\
    Europhysics Conf.\ on High Energy Physics}, Brussels, 1995. \\ 
    P.~Abreu \textit{et al.} (DELPHI), \zpc 66 (1995) 323. \\ 
    M. Acciarri \textit{et al.} (L3) \zpc 71 (1996) 379.\\ 
    R.~Akers \textit{et al.} (OPAL), \zpc 60 (1993) 199. 
\bibitem{n2} M.~Neubert, hep-ph/9610385, to appear in the proceedings
  of the \textit{20th John Hopkins Workshop on Current Problems in
         Particle Theory}, Heidelberg, 1996.
\bibitem{k} S.~Bertolini, F.~Borzumati and A.~Masiero, \npb 294 (1987)
         321.\\ 
         A.L.~Kagan, \prd 51 (1995) 6196.\\ 
         M.~Ciuchini, E.~Gabrielli and G.F.~Giudice, \plb 388 (1996)
         353.\\
         A.L.~Kagan and J.~Rathsman, hep-ph/9701300.  
\bibitem{ps} W.F.~Palmer and B.~Stech, \prd 48 (1993) 4174.
\bibitem{f} R.~Fleischer, \zpc\ 58 (1993) 483. 
\bibitem{acmp} G.~Altarelli, G.~Curci, G.~Martinelli and S.~Petrarca, 
               \npb 187 (1981) 461.
\bibitem{bbdm} W.~A.~Bardeen, A.~J.~Buras, D.~W.~Duke and T.~Muta,
               \prd 18 (1978) 3998.
\bibitem{flnn} A.F.~Falk, Z.~Ligeti, M.~Neubert and Y.~Nir, 
           \plb 326 (1994) 145.
\bibitem{b} G.~Buchalla, \npb 391 (1993) 501.
\bibitem{bf} A.~J.~Buras and R.~Fleischer, \plb 341 (1995) 379.
\bibitem{cfms} M.~Ciuchini, E.~Franco, G.~Martinelli and
         L. Silvestrini, hep-ph/9703353. 
\bibitem{cm} 
         M.A.~Shifman, A.I.~Vainshtein and V.I.~Zakharov, \prd 18 
         (1978) 2583 ; Erratum \textit{ibid.} D19 (1979) 2815.\\
         B.~Grinstein, R.~Springer and M.B.~Wise, \plb 202 (1988)
         138; \npb 339 (1990) 269. 
\bibitem{cfmrs} 
         M.~Ciuchini, E.~Franco, G.~Martinelli, L.~Reina and
         L.~Silvestrini, \plb 316 (1993) 127. \\ 
         M.~Ciuchini, E.~Franco, L.~Reina and
         L.~Silvestrini, \npb 421 (1994) 41.  \\
         M.~Ciuchini, E.~Franco, G.~Martinelli, L.~Reina and
         L.~Silvestrini, \plb 334 (1994) 137.  
\bibitem{bjlw} A.\ J.\ Buras, M.\ Jamin, M.\ E.\ Lautenbacher and 
              P.\ H.\ Weisz, \\ Nucl.\ Phys.\ B370 (1992) 69; 
              Addendum \textit{ibid} B375 (1992) 501.\\
              A.\ J.\ Buras, M.\ Jamin, M.\ E.\ Lautenbacher and 
              P.\ H.\ Weisz, \\ Nucl.\ Phys.\ B400 (1993) 37. 
\bibitem{beth} S.~Bethke, hep-ex/9609014, talk given at the 
  \textit{International Euroconference on Quantum Chromodynamics (QCD 96)},
  Montpellier, 1996. 
\bibitem{hn}  S.~Herrlich and U.~Nierste, \npb 455 (1995) 39. \\
              S.~Herrlich and U.~Nierste, \npb 476 (1996) 27.
\bibitem{sew} H.~Simma, G.~Eilam and D.~Wyler, \npb 352 (1991) 367.  
\bibitem{nir} Y.~Nir, \plb 221 (1989) 184.
\bibitem{bsu} I.~Bigi, M.~Shifman and N.~Uraltsev, hep-ph/9703290.
\bibitem{a} M.~Artuso, Nucl.~Instrum.~Meth.~A384 (1996) 39.
\bibitem{hn3}  S.~Herrlich and U.~Nierste, \prd  52 (1995) 6505.\\
  U.~Nierste,  hep-ph/9609310, 
  talk given at the \textit{Workshop on K physics}, Orsay, May/June
  1996. 


\end{thebibliography}
\end{document}